\documentclass[aps,prd,twocolumn,superscriptaddress,preprintnumbers,nofootinbib]{revtex4-2}
\usepackage{amsmath}
\usepackage{amssymb}
\usepackage{natbib}
\usepackage{graphicx}
\usepackage{epsf}
\usepackage{subfigure}
\usepackage{colortbl}
\usepackage[dvipsnames, table]{xcolor}
\usepackage{threeparttable}
\usepackage{dcolumn}
\usepackage{comment}
\usepackage{epsfig}
\usepackage{xspace}
\usepackage{multirow}
\usepackage{makecell}
\usepackage{mathtools}
\usepackage{appendix}
\usepackage{booktabs}
\usepackage[utf8]{inputenc}
\usepackage[TS1,T1]{fontenc}
\usepackage{diagbox}
\usepackage{ulem}
\usepackage[
colorlinks=True,
allcolors=blue,
linktocpage=true,
]{hyperref}
\usepackage[capitalise]{cleveref}
\usepackage{latexsym}
\usepackage{array}
\usepackage{orcidlink}
\usepackage[modulo,displaymath]{lineno}%
\usepackage{listings}
\usepackage{float}
\newcolumntype{C}[1]{>{\centering\let\newline\\\arraybackslash\hspace{0pt}}m{#1}}
\definecolor{background}{rgb}{0.95,0.95,0.92}
\definecolor{codeblack}{rgb}{0.1,0.1,0.1}
\definecolor{codegrey}{rgb}{0.3,0.3,0.3}
\definecolor{codered}{rgb}{0.6,0.1,0.1}
\definecolor{codegreen}{rgb}{0.1,0.6,0.1}
\definecolor{codeblue}{rgb}{0.1,0.1,0.6}

\newcommand\YAMLcolonstyle{\color{codered}\mdseries}
\newcommand\YAMLkeystyle{\color{codeblack}\bfseries}
\newcommand\YAMLvaluestyle{\color{codeblue}\mdseries}

\makeatletter

\newcommand\language@yaml{yaml}

\expandafter\expandafter\expandafter\lstdefinelanguage
\expandafter{\language@yaml}
{
  basicstyle=\YAMLkeystyle\ttfamily\footnotesize,
  keywords={true,false,null,y,n},
  keywordstyle=\color{codeblack}\bfseries,
  sensitive=false,
  comment=[l]{\#},
  morecomment=[s]{/*}{*/},
  commentstyle=\color{codegreen},
  stringstyle=\YAMLvaluestyle,
  moredelim=[l][\color{codered}]{\&},
  moredelim=[l][\color{codeblue}]{*},
  moredelim=**[il][\YAMLcolonstyle{:}\YAMLvaluestyle]{:},   % switch to value style at :
  morestring=[b]',
  morestring=[b]",
  literate =    {---}{{\ProcessThreeDashes}}3
                {>}{{\textcolor{codered}\textgreater}}1     
                {|}{{\textcolor{codered}\textbar}}1 
                {\ -\ }{{\mdseries\ -\ }}3,
  backgroundcolor=\color{background},
  numberstyle=\tiny\ttfamily\color{codegrey},
  numbers=left,
  numbersep=5pt,
  breaklines=true
}

\lst@AddToHook{EveryLine}{\ifx\lst@language\language@yaml\YAMLkeystyle\fi}
\makeatother

\newcommand\ProcessThreeDashes{\llap{\color{cyan}\mdseries-{-}-}}

\renewcommand{\arraystretch}{1.2}

\date{\today}

\DeclareGraphicsExtensions{.jpg,.pdf,.png,.eps,.ps}

\newcommand*{\lcdm}{$\Lambda$CDM}
\newcommand*{\planck}{\textit{Planck}}

\newcommand{\DESI}{\textsc{DESI}}

\newcommand{\ACT}{\textsc{ACT}} 
\newcommand{\ACTDR}{\textsc{ACT-DR6}} 

\newcommand{\ground}{\textsc{SPT+ACT}}
\newcommand{\cmball}{\textsc{CMB-SPA}}
\newcommand{\pact}{\textsc{P-ACT}}

\newcommand{\LCDM}{\lcdm}
\newcommand{\EDE}{AEDE}
\newcommand{\fede}{$f_{\rm EDE}$}
\newcommand{\DMAP}{$Q_{\rm DMAP}$}
\newcommand{\MPCL}{$Q_{\rm MPCL}$}
\newcommand{\AIC}{$\Delta$AIC}

\newcommand{\OLE}{\texttt{OL\'E}}

\newcommand{\RNum}[1]{\uppercase\expandafter{\romannumeral #1\relax}}

\definecolor{amber}{rgb}{1.0, 0.49, 0.0}

\definecolor{dblue}{RGB}{26, 158, 145}
\newcommand{\clipy}{\texttt{clipy}}

\def\sptlr{\text{SPT-3G D1}\xspace}

\def\cobaya{\texttt{Cobaya}\xspace}

\newcommand{\commenter}[1]{{}}

\begin{document}
\title{\sptlr: Axion Early Dark Energy with CMB experiments and DESI}
\author{A.~R.~Khalife\,\orcidlink{0000-0002-8388-4950}}
\affiliation{Sorbonne Universit\'e, CNRS, UMR 7095, Institut d'Astrophysique de Paris, 98 bis bd Arago, 75014 Paris, France}
\author{L.~Balkenhol\,\orcidlink{0000-0001-6899-1873}}
\affiliation{Sorbonne Universit\'e, CNRS, UMR 7095, Institut d'Astrophysique de Paris, 98 bis bd Arago, 75014 Paris, France}
\author{E.~Camphuis\,\orcidlink{0000-0003-3483-8461}}
\affiliation{Sorbonne Universit\'e, CNRS, UMR 7095, Institut d'Astrophysique de Paris, 98 bis bd Arago, 75014 Paris, France}
\author{A.~J.~Anderson\,\orcidlink{0000-0002-4435-4623}}
\affiliation{Fermi National Accelerator Laboratory, MS209, P.O. Box 500, Batavia, IL, 60510, USA}
\affiliation{Kavli Institute for Cosmological Physics, University of Chicago, 5640 South Ellis Avenue, Chicago, IL, 60637, USA}
\affiliation{Department of Astronomy and Astrophysics, University of Chicago, 5640 South Ellis Avenue, Chicago, IL, 60637, USA}
\author{B.~Ansarinejad}
\affiliation{School of Physics, University of Melbourne, Parkville, VIC 3010, Australia}
\author{M.~Archipley\,\orcidlink{0000-0002-0517-9842}}
\affiliation{Department of Astronomy and Astrophysics, University of Chicago, 5640 South Ellis Avenue, Chicago, IL, 60637, USA}
\affiliation{Kavli Institute for Cosmological Physics, University of Chicago, 5640 South Ellis Avenue, Chicago, IL, 60637, USA}
\author{P.~S.~Barry\,\orcidlink{0000-0001-9103-9354}}
\affiliation{School of Physics and Astronomy, Cardiff University, Cardiff, CF24 3AA, UK}
\author{K.~Benabed}
\affiliation{Sorbonne Universit\'e, CNRS, UMR 7095, Institut d'Astrophysique de Paris, 98 bis bd Arago, 75014 Paris, France}
\author{A.~N.~Bender\,\orcidlink{0000-0001-5868-0748}}
\affiliation{High-Energy Physics Division, Argonne National Laboratory, 9700 South Cass Avenue, Lemont, IL, 60439, USA}
\affiliation{Kavli Institute for Cosmological Physics, University of Chicago, 5640 South Ellis Avenue, Chicago, IL, 60637, USA}
\affiliation{Department of Astronomy and Astrophysics, University of Chicago, 5640 South Ellis Avenue, Chicago, IL, 60637, USA}
\author{B.~A.~Benson\,\orcidlink{0000-0002-5108-6823}}
\affiliation{Fermi National Accelerator Laboratory, MS209, P.O. Box 500, Batavia, IL, 60510, USA}
\affiliation{Kavli Institute for Cosmological Physics, University of Chicago, 5640 South Ellis Avenue, Chicago, IL, 60637, USA}
\affiliation{Department of Astronomy and Astrophysics, University of Chicago, 5640 South Ellis Avenue, Chicago, IL, 60637, USA}
\author{F.~Bianchini\,\orcidlink{0000-0003-4847-3483}}
\affiliation{Kavli Institute for Particle Astrophysics and Cosmology, Stanford University, 452 Lomita Mall, Stanford, CA, 94305, USA}
\affiliation{Department of Physics, Stanford University, 382 Via Pueblo Mall, Stanford, CA, 94305, USA}
\affiliation{SLAC National Accelerator Laboratory, 2575 Sand Hill Road, Menlo Park, CA, 94025, USA}
\author{L.~E.~Bleem\,\orcidlink{0000-0001-7665-5079}}
\affiliation{High-Energy Physics Division, Argonne National Laboratory, 9700 South Cass Avenue, Lemont, IL, 60439, USA}
\affiliation{Kavli Institute for Cosmological Physics, University of Chicago, 5640 South Ellis Avenue, Chicago, IL, 60637, USA}
\affiliation{Department of Astronomy and Astrophysics, University of Chicago, 5640 South Ellis Avenue, Chicago, IL, 60637, USA}
\author{F.~R.~Bouchet\,\orcidlink{0000-0002-8051-2924}}
\affiliation{Sorbonne Universit\'e, CNRS, UMR 7095, Institut d'Astrophysique de Paris, 98 bis bd Arago, 75014 Paris, France}
\author{L.~Bryant}
\affiliation{Enrico Fermi Institute, University of Chicago, 5640 South Ellis Avenue, Chicago, IL, 60637, USA}
\author{M.~G.~Campitiello}
\affiliation{High-Energy Physics Division, Argonne National Laboratory, 9700 South Cass Avenue, Lemont, IL, 60439, USA}
\author{J.~E.~Carlstrom\,\orcidlink{0000-0002-2044-7665}}
\affiliation{Kavli Institute for Cosmological Physics, University of Chicago, 5640 South Ellis Avenue, Chicago, IL, 60637, USA}
\affiliation{Enrico Fermi Institute, University of Chicago, 5640 South Ellis Avenue, Chicago, IL, 60637, USA}
\affiliation{Department of Physics, University of Chicago, 5640 South Ellis Avenue, Chicago, IL, 60637, USA}
\affiliation{High-Energy Physics Division, Argonne National Laboratory, 9700 South Cass Avenue, Lemont, IL, 60439, USA}
\affiliation{Department of Astronomy and Astrophysics, University of Chicago, 5640 South Ellis Avenue, Chicago, IL, 60637, USA}
\author{C.~L.~Chang}
\affiliation{High-Energy Physics Division, Argonne National Laboratory, 9700 South Cass Avenue, Lemont, IL, 60439, USA}
\affiliation{Kavli Institute for Cosmological Physics, University of Chicago, 5640 South Ellis Avenue, Chicago, IL, 60637, USA}
\affiliation{Department of Astronomy and Astrophysics, University of Chicago, 5640 South Ellis Avenue, Chicago, IL, 60637, USA}
\author{P.~Chaubal}
\affiliation{School of Physics, University of Melbourne, Parkville, VIC 3010, Australia}
\author{P.~M.~Chichura\,\orcidlink{0000-0002-5397-9035}}
\affiliation{Department of Physics, University of Chicago, 5640 South Ellis Avenue, Chicago, IL, 60637, USA}
\affiliation{Kavli Institute for Cosmological Physics, University of Chicago, 5640 South Ellis Avenue, Chicago, IL, 60637, USA}
\author{A.~Chokshi}
\affiliation{Department of Astronomy and Astrophysics, University of Chicago, 5640 South Ellis Avenue, Chicago, IL, 60637, USA}
\author{T.-L.~Chou\,\orcidlink{0000-0002-3091-8790}}
\affiliation{Department of Astronomy and Astrophysics, University of Chicago, 5640 South Ellis Avenue, Chicago, IL, 60637, USA}
\affiliation{Kavli Institute for Cosmological Physics, University of Chicago, 5640 South Ellis Avenue, Chicago, IL, 60637, USA}
\affiliation{National Taiwan University, No. 1, Sec. 4, Roosevelt Road, Taipei 106319, Taiwan}
\author{A.~Coerver}
\affiliation{Department of Physics, University of California, Berkeley, CA, 94720, USA}
\author{T.~M.~Crawford\,\orcidlink{0000-0001-9000-5013}}
\affiliation{Department of Astronomy and Astrophysics, University of Chicago, 5640 South Ellis Avenue, Chicago, IL, 60637, USA}
\affiliation{Kavli Institute for Cosmological Physics, University of Chicago, 5640 South Ellis Avenue, Chicago, IL, 60637, USA}
\author{C.~Daley\,\orcidlink{0000-0002-3760-2086}}
\affiliation{Universit\'e Paris-Saclay, Universit\'e Paris Cit\'e, CEA, CNRS, AIM, 91191, Gif-sur-Yvette, France}
\affiliation{Department of Astronomy, University of Illinois Urbana-Champaign, 1002 West Green Street, Urbana, IL, 61801, USA}
\author{T.~de~Haan}
\affiliation{High Energy Accelerator Research Organization (KEK), Tsukuba, Ibaraki 305-0801, Japan}
\author{K.~R.~Dibert}
\affiliation{Department of Astronomy and Astrophysics, University of Chicago, 5640 South Ellis Avenue, Chicago, IL, 60637, USA}
\affiliation{Kavli Institute for Cosmological Physics, University of Chicago, 5640 South Ellis Avenue, Chicago, IL, 60637, USA}
\author{M.~A.~Dobbs}
\affiliation{Department of Physics and McGill Space Institute, McGill University, 3600 Rue University, Montreal, Quebec H3A 2T8, Canada}
\affiliation{Canadian Institute for Advanced Research, CIFAR Program in Gravity and the Extreme Universe, Toronto, ON, M5G 1Z8, Canada}
\author{M.~Doohan}
\affiliation{School of Physics, University of Melbourne, Parkville, VIC 3010, Australia}
\author{A.~Doussot}
\affiliation{Sorbonne Universit\'e, CNRS, UMR 7095, Institut d'Astrophysique de Paris, 98 bis bd Arago, 75014 Paris, France}
\author{D.~Dutcher\,\orcidlink{0000-0002-9962-2058}}
\affiliation{Joseph Henry Laboratories of Physics, Jadwin Hall, Princeton University, Princeton, NJ 08544, USA}
\author{W.~Everett}
\affiliation{Department of Astrophysical and Planetary Sciences, University of Colorado, Boulder, CO, 80309, USA}
\author{C.~Feng}
\affiliation{Department of Physics, University of Illinois Urbana-Champaign, 1110 West Green Street, Urbana, IL, 61801, USA}
\author{K.~R.~Ferguson\,\orcidlink{0000-0002-4928-8813}}
\affiliation{Department of Physics and Astronomy, University of California, Los Angeles, CA, 90095, USA}
\affiliation{Department of Physics and Astronomy, Michigan State University, East Lansing, MI 48824, USA}
\author{K.~Fichman}
\affiliation{Department of Physics, University of Chicago, 5640 South Ellis Avenue, Chicago, IL, 60637, USA}
\affiliation{Kavli Institute for Cosmological Physics, University of Chicago, 5640 South Ellis Avenue, Chicago, IL, 60637, USA}
\author{A.~Foster\,\orcidlink{0000-0002-7145-1824}}
\affiliation{Joseph Henry Laboratories of Physics, Jadwin Hall, Princeton University, Princeton, NJ 08544, USA}
\author{S.~Galli}
\affiliation{Sorbonne Universit\'e, CNRS, UMR 7095, Institut d'Astrophysique de Paris, 98 bis bd Arago, 75014 Paris, France}
\author{A.~E.~Gambrel}
\affiliation{Kavli Institute for Cosmological Physics, University of Chicago, 5640 South Ellis Avenue, Chicago, IL, 60637, USA}
\author{R.~W.~Gardner}
\affiliation{Enrico Fermi Institute, University of Chicago, 5640 South Ellis Avenue, Chicago, IL, 60637, USA}
\author{F.~Ge}
\affiliation{Kavli Institute for Particle Astrophysics and Cosmology, Stanford University, 452 Lomita Mall, Stanford, CA, 94305, USA}
\affiliation{Department of Physics, Stanford University, 382 Via Pueblo Mall, Stanford, CA, 94305, USA}
\affiliation{Department of Physics \& Astronomy, University of California, One Shields Avenue, Davis, CA 95616, USA}
\author{N.~Goeckner-Wald}
\affiliation{Department of Physics, Stanford University, 382 Via Pueblo Mall, Stanford, CA, 94305, USA}
\affiliation{Kavli Institute for Particle Astrophysics and Cosmology, Stanford University, 452 Lomita Mall, Stanford, CA, 94305, USA}
\author{R.~Gualtieri\,\orcidlink{0000-0003-4245-2315}}
\affiliation{High-Energy Physics Division, Argonne National Laboratory, 9700 South Cass Avenue, Lemont, IL, 60439, USA}
\affiliation{Department of Physics and Astronomy, Northwestern University, 633 Clark St, Evanston, IL, 60208, USA}
\author{F.~Guidi\,\orcidlink{0000-0001-7593-3962}}
\affiliation{Sorbonne Universit\'e, CNRS, UMR 7095, Institut d'Astrophysique de Paris, 98 bis bd Arago, 75014 Paris, France}
\author{S.~Guns}
\affiliation{Department of Physics, University of California, Berkeley, CA, 94720, USA}
\author{N.~W.~Halverson}
\affiliation{CASA, Department of Astrophysical and Planetary Sciences, University of Colorado, Boulder, CO, 80309, USA }
\affiliation{Department of Physics, University of Colorado, Boulder, CO, 80309, USA}
\author{E.~Hivon\,\orcidlink{0000-0003-1880-2733}}
\affiliation{Sorbonne Universit\'e, CNRS, UMR 7095, Institut d'Astrophysique de Paris, 98 bis bd Arago, 75014 Paris, France}
% \author{G.~P.~Holder\,\orcidlink{0000-0002-0463-6394}}
% \affiliation{Department of Physics, University of Illinois Urbana-Champaign, 1110 West Green Street, Urbana, IL, 61801, USA}
\author{W.~L.~Holzapfel}
\affiliation{Department of Physics, University of California, Berkeley, CA, 94720, USA}
\author{J.~C.~Hood}
\affiliation{Kavli Institute for Cosmological Physics, University of Chicago, 5640 South Ellis Avenue, Chicago, IL, 60637, USA}
\author{A.~Hryciuk}
\affiliation{Department of Physics, University of Chicago, 5640 South Ellis Avenue, Chicago, IL, 60637, USA}
\affiliation{Kavli Institute for Cosmological Physics, University of Chicago, 5640 South Ellis Avenue, Chicago, IL, 60637, USA}
\author{N.~Huang\,\orcidlink{0000-0003-3595-0359}}
\affiliation{Department of Physics, University of California, Berkeley, CA, 94720, USA}
\author{F.~K\'eruzor\'e}
\affiliation{High-Energy Physics Division, Argonne National Laboratory, 9700 South Cass Avenue, Lemont, IL, 60439, USA}
\author{L.~Knox}
\affiliation{Department of Physics \& Astronomy, University of California, One Shields Avenue, Davis, CA 95616, USA}
\author{M.~Korman}
\affiliation{Department of Physics, Case Western Reserve University, Cleveland, OH, 44106, USA}
\author{K.~Kornoelje}
\affiliation{Department of Astronomy and Astrophysics, University of Chicago, 5640 South Ellis Avenue, Chicago, IL, 60637, USA}
\affiliation{Kavli Institute for Cosmological Physics, University of Chicago, 5640 South Ellis Avenue, Chicago, IL, 60637, USA}
\affiliation{High-Energy Physics Division, Argonne National Laboratory, 9700 South Cass Avenue, Lemont, IL, 60439, USA}
\author{C.-L.~Kuo}
\affiliation{Kavli Institute for Particle Astrophysics and Cosmology, Stanford University, 452 Lomita Mall, Stanford, CA, 94305, USA}
\affiliation{Department of Physics, Stanford University, 382 Via Pueblo Mall, Stanford, CA, 94305, USA}
\affiliation{SLAC National Accelerator Laboratory, 2575 Sand Hill Road, Menlo Park, CA, 94025, USA}
\author{K.~Levy}
\affiliation{School of Physics, University of Melbourne, Parkville, VIC 3010, Australia}
\author{A.~E.~Lowitz\,\orcidlink{0000-0002-4747-4276}}
\affiliation{Kavli Institute for Cosmological Physics, University of Chicago, 5640 South Ellis Avenue, Chicago, IL, 60637, USA}
\author{C.~Lu}
\affiliation{Department of Physics, University of Illinois Urbana-Champaign, 1110 West Green Street, Urbana, IL, 61801, USA}
\author{G.~P.~Lynch\,\orcidlink{0009-0004-3143-1708}}
\affiliation{Department of Physics \& Astronomy, University of California, One Shields Avenue, Davis, CA 95616, USA}
\author{A.~Maniyar}
\affiliation{Kavli Institute for Particle Astrophysics and Cosmology, Stanford University, 452 Lomita Mall, Stanford, CA, 94305, USA}
\affiliation{Department of Physics, Stanford University, 382 Via Pueblo Mall, Stanford, CA, 94305, USA}
\affiliation{SLAC National Accelerator Laboratory, 2575 Sand Hill Road, Menlo Park, CA, 94025, USA}
\author{E.~S.~Martsen}
\affiliation{Department of Astronomy and Astrophysics, University of Chicago, 5640 South Ellis Avenue, Chicago, IL, 60637, USA}
\affiliation{Kavli Institute for Cosmological Physics, University of Chicago, 5640 South Ellis Avenue, Chicago, IL, 60637, USA}
\author{F.~Menanteau}
\affiliation{Department of Astronomy, University of Illinois Urbana-Champaign, 1002 West Green Street, Urbana, IL, 61801, USA}
\affiliation{Center for AstroPhysical Surveys, National Center for Supercomputing Applications, Urbana, IL, 61801, USA}
\author{M.~Millea\,\orcidlink{0000-0001-7317-0551}}
\affiliation{Department of Physics, University of California, Berkeley, CA, 94720, USA}
\author{J.~Montgomery}
\affiliation{Department of Physics and McGill Space Institute, McGill University, 3600 Rue University, Montreal, Quebec H3A 2T8, Canada}
\author{Y.~Nakato}
\affiliation{Department of Physics, Stanford University, 382 Via Pueblo Mall, Stanford, CA, 94305, USA}
\author{T.~Natoli}
\affiliation{Kavli Institute for Cosmological Physics, University of Chicago, 5640 South Ellis Avenue, Chicago, IL, 60637, USA}
\author{G.~I.~Noble\,\orcidlink{0000-0002-5254-243X}}
\affiliation{Dunlap Institute for Astronomy \& Astrophysics, University of Toronto, 50 St. George Street, Toronto, ON, M5S 3H4, Canada}
\affiliation{David A. Dunlap Department of Astronomy \& Astrophysics, University of Toronto, 50 St. George Street, Toronto, ON, M5S 3H4, Canada}
\author{Y.~Omori}
\affiliation{Department of Astronomy and Astrophysics, University of Chicago, 5640 South Ellis Avenue, Chicago, IL, 60637, USA}
\affiliation{Kavli Institute for Cosmological Physics, University of Chicago, 5640 South Ellis Avenue, Chicago, IL, 60637, USA}
\author{A.~Ouellette}
\affiliation{Department of Physics, University of Illinois Urbana-Champaign, 1110 West Green Street, Urbana, IL, 61801, USA}
\author{Z.~Pan\,\orcidlink{0000-0002-6164-9861}}
\affiliation{High-Energy Physics Division, Argonne National Laboratory, 9700 South Cass Avenue, Lemont, IL, 60439, USA}
\affiliation{Kavli Institute for Cosmological Physics, University of Chicago, 5640 South Ellis Avenue, Chicago, IL, 60637, USA}
\affiliation{Department of Physics, University of Chicago, 5640 South Ellis Avenue, Chicago, IL, 60637, USA}
\author{P.~Paschos}
\affiliation{Enrico Fermi Institute, University of Chicago, 5640 South Ellis Avenue, Chicago, IL, 60637, USA}
\author{K.~A.~Phadke\,\orcidlink{0000-0001-7946-557X}}
\affiliation{Department of Astronomy, University of Illinois Urbana-Champaign, 1002 West Green Street, Urbana, IL, 61801, USA}
\affiliation{Center for AstroPhysical Surveys, National Center for Supercomputing Applications, Urbana, IL, 61801, USA}
\affiliation{NSF-Simons AI Institute for the Sky (SkAI), 172 E. Chestnut St., Chicago, IL 60611, USA}
\author{A.~W.~Pollak}
\affiliation{Department of Astronomy and Astrophysics, University of Chicago, 5640 South Ellis Avenue, Chicago, IL, 60637, USA}
\author{K.~Prabhu}
\affiliation{Department of Physics \& Astronomy, University of California, One Shields Avenue, Davis, CA 95616, USA}
\author{W.~Quan}
\affiliation{High-Energy Physics Division, Argonne National Laboratory, 9700 South Cass Avenue, Lemont, IL, 60439, USA}
\affiliation{Department of Physics, University of Chicago, 5640 South Ellis Avenue, Chicago, IL, 60637, USA}
\affiliation{Kavli Institute for Cosmological Physics, University of Chicago, 5640 South Ellis Avenue, Chicago, IL, 60637, USA}
% \author{S.~Raghunathan\,\orcidlink{0000-0003-1405-378X}}
% \affiliation{Center for AstroPhysical Surveys, National Center for Supercomputing Applications, Urbana, IL, 61801, USA}
\author{M.~Rahimi}
\affiliation{School of Physics, University of Melbourne, Parkville, VIC 3010, Australia}
\author{A.~Rahlin\,\orcidlink{0000-0003-3953-1776}}
\affiliation{Department of Astronomy and Astrophysics, University of Chicago, 5640 South Ellis Avenue, Chicago, IL, 60637, USA}
\affiliation{Kavli Institute for Cosmological Physics, University of Chicago, 5640 South Ellis Avenue, Chicago, IL, 60637, USA}
\author{C.~L.~Reichardt\,\orcidlink{0000-0003-2226-9169}}
\affiliation{School of Physics, University of Melbourne, Parkville, VIC 3010, Australia}
\author{M.~Rouble}
\affiliation{Department of Physics and McGill Space Institute, McGill University, 3600 Rue University, Montreal, Quebec H3A 2T8, Canada}
\author{J.~E.~Ruhl}
\affiliation{Department of Physics, Case Western Reserve University, Cleveland, OH, 44106, USA}
\author{E.~Schiappucci}
\affiliation{School of Physics, University of Melbourne, Parkville, VIC 3010, Australia}
\author{A.~Simpson}
\affiliation{Department of Astronomy and Astrophysics, University of Chicago, 5640 South Ellis Avenue, Chicago, IL, 60637, USA}
\affiliation{Kavli Institute for Cosmological Physics, University of Chicago, 5640 South Ellis Avenue, Chicago, IL, 60637, USA}
\author{J.~A.~Sobrin\,\orcidlink{0000-0001-6155-5315}}
\affiliation{Fermi National Accelerator Laboratory, MS209, P.O. Box 500, Batavia, IL, 60510, USA}
\affiliation{Kavli Institute for Cosmological Physics, University of Chicago, 5640 South Ellis Avenue, Chicago, IL, 60637, USA}
\author{A.~A.~Stark}
\affiliation{Center for Astrophysics \textbar{} Harvard \& Smithsonian, 60 Garden Street, Cambridge, MA, 02138, USA}
\author{J.~Stephen}
\affiliation{Enrico Fermi Institute, University of Chicago, 5640 South Ellis Avenue, Chicago, IL, 60637, USA}
\author{C.~Tandoi}
\affiliation{Department of Astronomy, University of Illinois Urbana-Champaign, 1002 West Green Street, Urbana, IL, 61801, USA}
\author{B.~Thorne}
\affiliation{Department of Physics \& Astronomy, University of California, One Shields Avenue, Davis, CA 95616, USA}
\author{C.~Trendafilova}
\affiliation{Center for AstroPhysical Surveys, National Center for Supercomputing Applications, Urbana, IL, 61801, USA}
\author{C.~Umilta\,\orcidlink{0000-0002-6805-6188}}
\affiliation{Department of Physics, University of Illinois Urbana-Champaign, 1110 West Green Street, Urbana, IL, 61801, USA}
\author{J.~D.~Vieira\,\orcidlink{0000-0001-7192-3871}}
\affiliation{Department of Astronomy, University of Illinois Urbana-Champaign, 1002 West Green Street, Urbana, IL, 61801, USA}
\affiliation{Department of Physics, University of Illinois Urbana-Champaign, 1110 West Green Street, Urbana, IL, 61801, USA}
\affiliation{Center for AstroPhysical Surveys, National Center for Supercomputing Applications, Urbana, IL, 61801, USA}
\author{A.~Vitrier\,\orcidlink{0009-0009-3168-092X}}
\affiliation{Sorbonne Universit\'e, CNRS, UMR 7095, Institut d'Astrophysique de Paris, 98 bis bd Arago, 75014 Paris, France}
\author{Y.~Wan}
\affiliation{Department of Astronomy, University of Illinois Urbana-Champaign, 1002 West Green Street, Urbana, IL, 61801, USA}
\affiliation{Center for AstroPhysical Surveys, National Center for Supercomputing Applications, Urbana, IL, 61801, USA}
\author{N.~Whitehorn\,\orcidlink{0000-0002-3157-0407}}
\affiliation{Department of Physics and Astronomy, Michigan State University, East Lansing, MI 48824, USA}
\author{W.~L.~K.~Wu\,\orcidlink{0000-0001-5411-6920}}
\affiliation{Kavli Institute for Particle Astrophysics and Cosmology, Stanford University, 452 Lomita Mall, Stanford, CA, 94305, USA}
\affiliation{SLAC National Accelerator Laboratory, 2575 Sand Hill Road, Menlo Park, CA, 94025, USA}
\author{M.~R.~Young}
\affiliation{Fermi National Accelerator Laboratory, MS209, P.O. Box 500, Batavia, IL, 60510, USA}
\affiliation{Kavli Institute for Cosmological Physics, University of Chicago, 5640 South Ellis Avenue, Chicago, IL, 60637, USA}
\author{J.~A.~Zebrowski}
\affiliation{Kavli Institute for Cosmological Physics, University of Chicago, 5640 South Ellis Avenue, Chicago, IL, 60637, USA}
\affiliation{Department of Astronomy and Astrophysics, University of Chicago, 5640 South Ellis Avenue, Chicago, IL, 60637, USA}
\affiliation{Fermi National Accelerator Laboratory, MS209, P.O. Box 500, Batavia, IL, 60510, USA}
\collaboration{SPT-3G Collaboration}
\noaffiliation

\begin{abstract}
We present the most up-to-date constraints on axion early dark energy (AEDE) from cosmic microwave background (CMB) and baryon acoustic oscillation (BAO) measurements. In particular, we assess the impact of data from ground-based CMB experiments, the South Pole Telescope (SPT) and the Atacama Cosmology Telescope (\ACT{})---both with and without \planck{}---on constraints on \EDE{}. We also highlight the impact that BAO information from the Dark Energy Spectroscopic Instrument (\DESI{}) has on these constraints. From CMB data alone, we do not find statistically significant evidence for the presence of \EDE{}, and we find only moderate reduction in the Hubble tension. From the latest SPT data alone, we find the maximal fractional contribution of AEDE to the cosmic energy budget is $f_{\rm EDE}\,<\,0.12$ at 95\% confidence level (CL), and the Hubble tension between the SPT and SH0ES results is reduced to the $2.3\,\sigma$ level. When combining the latest SPT, ACT, and \planck{} datasets, we find $f_{\rm EDE}\,<\,0.070$ at 95\% CL and the Hubble tension at the $3.6\, \sigma$ level. In contrast, adding \DESI{} data to the CMB datasets results in mild preference for AEDE and, in some cases, non-negligible reduction in the Hubble tension. From SPT+DESI, we find $f_{\rm EDE}\,=\,0.081^{+0.037}_{-0.052}$ at 68\% CL, and the Hubble tension reduces to $1.5\,\sigma$. From the combination of \DESI{} with all three CMB experiments, we get $f_{\rm EDE}\,=\, 0.055^{+0.024}_{-0.047}$ at 68\% CL and a weak preference for \EDE{} over \LCDM{}. This data combination, in turn, reduces the Hubble tension to $2.6\, \sigma$. We highlight that this shift in parameters when adding the \DESI{} dataset is a manifestation of the discrepancy currently present between \DESI{} and CMB experiments in the concordance model \LCDM{}.
\end{abstract}
\maketitle
\flushbottom

\section{Introduction}
In recent months, three major datasets have been released that provide significant new constraints on cosmological parameters. These are the South Pole Telescope's third generation camera's (SPT-3G) observations of the main field taken during 2019 and 2020 (\sptlr{})~\cite{Camphuis_etal,MUSE,SPT_Maps_Paper}
%~\citep[W. Quan et al., in preparation]{Camphuis_etal,MUSE}
, the Atacama Cosmology Telescope's sixth data release (ACT-DR6)~\cite{ACTDR6_Maps,ACT_DR6_1,ACTDR6_Extended}, and the second release of the Dark Energy Spectroscopic Instrument (DESI-DR2)~\cite{desi2025,DESI}. \sptlr{} is the first in a series of SPT releases; following releases will improve the constraints even further~\cite{Prabhu_etal}. With this trio of datasets, the time is auspicious to revisit one of the most promising models to ease the Hubble tension: axion early dark energy (\EDE{})~\cite{EDE3,EDE2,EDE1,Ups_Downs_EDE}.

%As a reminder, the Hubble tension is the growing discrepancy in the determination of the expansion rate today, $H_0$, between early Universe probes and the ``Supernovae, H0, for the Equation of State of Dark energy'' (SH0ES) collaboration. The latter has presented its most recent measurement of $H_0$ to be $73.17 \pm 0.86$ km/s/Mpc~\cite{Breuval:2024lsv}. On the other hand, assuming the concordance model of cosmology $\Lambda$CDM,~\cite{Camphuis_etal} reported $H_0 = 67.23 \pm 0.35$ km/s/Mpc from \sptlr{}, \ACTDR{} and \planck{} combined\footnote{Each dataset includes its corresponding lensing likelihood: ref.~\cite{MUSE} for SPT-3G,~\cite{ACT_DR6_lens} for \ACTDR{} and~\cite{PR4lens} for \planck{}.}, bringing the tension to the $6.4\, \sigma$ level.
As a reminder, the Hubble tension is a discrepancy between cosmological model-dependent inferences of the expansion rate today, $H_0$, and those using the classical distance ladder~\cite{Cosmology_intertwined,In_Realm,Camarena_2019,Camarena_2021,Verde1,Trouble_H0,VerdeTreuRiess,H0_Olympics}. The latter method more directly determines $H_0$ and is much less sensitive to cosmological model assumptions. The most precise model-dependent inference of $H_0$ comes from the CMB. Within \LCDM{}, the combination of \sptlr{}, \ACTDR{}, and \planck{} results in $H_0 = 67.19 \, \pm \, 0.38$ km/s/Mpc~\cite{Camphuis_etal}.\footnote{Each dataset contains $TT/TE/EE$ data and their corresponding lensing likelihood:~\cite{MUSE} for \sptlr{},~\cite{ACT_DR6_lens} for \ACTDR{}, and~\cite{PR4lens} for \planck{} (see~\cite{Camphuis_etal} for a detailed description of the datasets).} The most precise result using the classical distance ladder is  from the ``Supernovae, H0, for the Equation of State of Dark Energy'' (SH0ES) Collaboration, which finds $H_0 = 73.17 \, \pm \, 0.86$ km/s/Mpc~\cite{Breuval:2024lsv}. This value is discrepant with the CMB one mentioned above by $6.4\, \sigma$~\cite{Camphuis_etal}. 
The quest to solve this tension has resulted in a plentitude of theoretical models~\cite{Khalife:2023qbu,H0_Olympics}, of which \EDE{} is one of the most promising solutions. It is therefore interesting to update constraints on this model with the new datasets and see whether it still relieves the Hubble tension.

In this work, which is a follow-up to the cosmological constraints in~\cite{Camphuis_etal}, we present constraints on \EDE{} from the datasets mentioned above. We highlight the improvement in constraints due to the addition of \sptlr{} and \ACTDR{} to \planck{} data and contrast constraints coming from the CMB alone to those with BAO data from \DESI{}-DR2. In Section~\ref{Sec:EDE_Theory}, we present a concise description of the model, our assessment criteria for the ability of \EDE{} to solve the Hubble tension, and the numerical setup. We list the different datasets considered in this work in Section~\ref{Sec: Data_Sets}. This Section is followed by a presentation of our main results and their implications for the status of \EDE{} from CMB data alone (Section~\ref{Sec:CMB_alone}) and in combination with \DESI{}-DR2 BAO data (Section~\ref{Sec:CMB_and_BAO}). We end with some concluding remarks in Section~\ref{Sec:Conclusion}. 

\section{Early Dark Energy: A Brief Overview}
\label{Sec:EDE_Theory}

Early dark energy (EDE) models are
%In this section, we give a summarized description of EDE.
%This type of model is 
motivated by higher-dimensional theories (e.g. string theory) that predict the existence of scalar fields~\cite{AxiString} and have been considered in many works prior to the appearance of the Hubble tension~\cite{Old_EDE1,Old_EDE2,Old_EDE3}. At the background level, the presence of such scalar fields can add to the energy budget of the Universe prior to recombination, increasing the Hubble parameter at that epoch, $H(z)$. This in turn decreases the sound horizon in a way that could compensate for the increase in $H_0$ needed to solve the Hubble tension~\cite[e.g.][]{knox19}. Moreover, the presence of such a field has additional nontrivial impacts on the dynamics of the Universe. For instance, the amount of energy density carried by the field changes the damping scale of the CMB, the amplitude of the Sachs-Wolfe effect~\cite{Sachs-Wolfe} (including the early integrated one), the evolution of matter perturbations, and the ``radiation driving'' of acoustic oscillation amplitudes \cite{hu97}
(see~\cite{Ups_Downs_EDE} for more details).

Several EDE models have been proposed, each involving different mechanisms (see~\cite{EDE1,NEDE,Elisa-Eiishiro,Elisa-Laura,Birefrengence1,Birefrengence2,EDE_Curvature,Tristan_Nils} for more details). One particular model that captures a great deal of phenomenology with a fairly simple prescription is \EDE{}~\cite{EDE1,AxiCLASS1,AxiEDE_String1,AxiEDE_String2}, where an axion field ($\phi$) causes the above-mentioned effects of EDE, with a potential
\begin{equation}
	V(\phi) = m^2f^2\left[1-\cos(\theta)\right]^n,
\end{equation}
where $m$ is the field's mass, $f$ is its decay constant, $n$ is an integer, and $\theta = \phi/f$. The initial value of the latter, $\theta_{\rm i}$, is a free parameter of the model. Following previous works~\cite{Khalife:2023qbu,H0_Olympics,EDE_Silvia_Lennart,EDE_ACT1,EDE_ACT2,EDE_Curvature}, we consider the case\footnote{As shown in the works cited above, the case $n=3$ is the most promising in easing the tension. This is partially attributed to the fact that in this case $\phi$ dilutes faster than radiation after it becomes dynamical, as opposed to the case $n=2$ ($n=1$), where $\phi$ dilutes as radiation (matter)~\cite{EDE1,AxiCLASS1}.} $n=3$ and substitute $m$ and $f$ with the phenomenological parameters $z_c$ and $f_{\mathrm{EDE}}$~\cite{EDE1,AxiCLASS1}. The former is the critical redshift at which $\phi$ becomes dynamical and its energy density decays faster than that of radiation, while the latter,
\begin{equation}
    f_\mathrm{EDE} = \left. \frac{\rho_\mathrm{EDE}(z)}{3H(z)^2/8\pi G}\right|_{z=z_c},
\end{equation}
is the fraction of energy density occupied by the axion field at $z_c$.\footnote{In units where the reduced Planck constant, $\hbar$, and the speed of light, $c$, are 1. We are also considering a spatially flat universe.} To solve the Hubble tension, one would typically need $f_{\rm EDE} \sim 10\%$ with $z_c$ close to matter-radiation equality, i.e. $z_c\sim 10^3-10^4$~\cite{EDE1,Ups_Downs_EDE}.

To assess the ability of the model to solve the Hubble tension, we use three metrics: (1) marginalized posterior compatibility level, (2) difference of the maximum a posteriori, and (3) Akaike information criterion. 

\textbf{Marginalized Posterior Compatibility Level} (MPCL):
this metric, denoted hereafter as $Q_{\rm MPCL}$, is defined in~\cite{Khalife:2023qbu} (see~\cite{Q_MPCL1,Q_MPCL2} for more details). Briefly, this Bayesian statistic quantifies how much the inferred value of $H_0$ in \EDE{} with a given dataset deviates from the value measured by SH0ES without assuming Gaussianity of the posteriors. In other words, it is a generalization of the rule-of-thumb difference in mean introduced in~\cite{Concordance_Cosmology}. We consider $Q_{\rm MPCL}\leq 3\, \sigma$ (see~\cite{Khalife:2023qbu} for the meaning of $\sigma$) to be a passing score for the model. 

\textbf{Difference of the Maximum A Posteriori} (DMAP): this is a frequentist statistic that measures the change in the best-fit $\chi^2$ for a given data combination, within a model, due to the addition of information from SH0ES. The advantage of considering this metric along with \MPCL{} is twofold. First, it measures the ability of the whole model to fit the data, rather than focusing on only one parameter ($H_0$ in our case). Second, since it is a frequentist quantity, it avoids biases that could be due to the choice of priors (see~\cite{Khalife:2023qbu,H0_Olympics} for more details). This statistic is defined as~\cite{Concordance_Cosmology}
 \begin{equation}
     Q_{\rm DMAP} = \sqrt{\chi^2_{\rm w/\ SH0ES} - \chi^2_{\rm w/o\ SH0ES}},
     \label{eq:Q_DMAP}
 \end{equation}
 where $\chi^2_{\rm w/ \  SH0ES} (\chi^2_{\rm w/o \  SH0ES})$ corresponds to the minimum $\chi^2$ value of \EDE{} for a given dataset with (without) SH0ES information. A value of $Q_{\rm DMAP}\leq3\, \sigma$\footnote{The number of $\sigma$s for this metric corresponds to the Gaussian equivalent of a probability to exceed (PTE) from a $\Delta\chi^2$ for a $\chi^2$-distribution with 1 degree of freedom~\cite{Concordance_Cosmology,Efstathiou_Poulin_EDE}.} is considered a passing value for the model with a given dataset. 
 
 \textbf{Akaike Information Criterion} (AIC):
 also a frequentist statistic, this metric computes the improvement in fit a model has compared to \LCDM{} for a given dataset. Although it does not quantify the tension on $H_0$ with SH0ES, it gives important information when judging if the model is an acceptable solution to the tension. For the model at hand, it is defined as~\cite{AIC1,Bayesian1}:
 \begin{equation}
     \Delta{\rm AIC} = \chi^2_{\rm AEDE} - \chi^2_{\Lambda \rm CDM} + 2N,
     \label{eq:AIC}
 \end{equation}
where $\chi^2_{\rm AEDE}$ ($\chi^2_{\Lambda \rm CDM}$) is the minimum $\chi^2$ of \EDE{} (\LCDM{}) for a given dataset (which will be specified in Section~\ref{Sec:Results}) and $N$ is the additional number of parameters that \EDE{} has relative to \LCDM{} ($N=3$ in this case). To pass this metric, we require $\Delta{\rm AIC}\leq -6.91$, which corresponds to more than a ``weak preference'' on the Jeffreys scale~\cite{Jeffreys1939-JEFTOP-5,H0_Olympics,Bellido_Jeffrey}. Note that this metric has been used in previous works for datasets that include SH0ES~\cite{H0_Olympics,Khalife:2023qbu}. However, we focus here on the ability of the model to fit the data without SH0ES. This focuses us on what we think is the more interesting question: is there SH0ES-independent evidence for the \EDE{} model? 

If \EDE{} passes all three tests, we consider it as a viable solution to the Hubble tension. Although this set of metrics is not complete, it gives a reasonable indication of how well the model reduces the tension and fits the current datasets. 

However, it is important to note that this model suffers from prior volume effects~\cite{Elisa-Eiishiro,Tristan_Profile}. As $f_{\rm EDE}\rightarrow0$, the model effectively becomes \LCDM{}. When performing a Markov chain Monte Carlo (MCMC), this limit could be reached for a large range of values for the additional parameters ($\theta_{\rm i}$ and $z_c$). This will result in additional weight on that parameter space, and thus shift the samples' distribution of the MCMC toward the \LCDM{} region. Another complication arises due to the choice of the prior on $z_c$. If the chosen prior range is higher than the recombination redshift, this will result in higher values of \fede{} but without significant change in $H_0$~\cite{LaPosta:2021pgm}. Avoiding these prior-related effects motivates the use of the prior-independent, i.e. frequentist, statistics such as \DMAP{} or \AIC{} we use here. To determine a confidence interval that is not affected by this effect, one would need to perform a profile likelihood analysis and compute these intervals with techniques such as in~\cite{Feldman_Profile}. Such an analysis is beyond the scope of the current work, but it is considered in~\cite{Jhaveri:2026bla}.%[Jhaveri et al, in preparation].
\subsection{Numerical setup}
\label{Sec:Num_Set}
We use the implementation of the \EDE{} model in the Boltzmann code \texttt{AxiCLASS}~\cite{AxiCLASS1,AxiCLASS2}\footnote{\url{https://github.com/PoulinV/AxiCLASS}. Note that we used version \texttt{v3.2.0} of the code, with the precision parameters set as described in Appendix A of~\cite{ACTDR6_Extended}.}, a modified version of \texttt{CLASS}~\cite{CLASS1,CLASS2}. We impose uniform priors on the parameters of the model,
\begin{align}
\label{Eq:EDE_priors}
    &f_{\mathrm{EDE}}\sim\mathcal{U}(0,0.5), \nonumber \\ 
    &\theta_{\rm i}\sim\mathcal{U}(0.01,3.1), \ 
    \mathrm{and} \nonumber \\
    &\log_{10}(a_c)\sim\mathcal{U}(-4.5,-2.8),
\end{align}
% \begin{equation*}
% \label{Eq:EDE_priors}
%     f_{\mathrm{EDE}}\sim\mathcal{U}(0,0.5), \ \theta_{\rm i}\sim\mathcal{U}(0.01,3.1), \ 
%     \mathrm{and} \ \log_{10}(a_c)\sim\mathcal{U}(-4.5,-2.8),
% \end{equation*}
where $a_c$ is the scale factor at $z_c$ and $\mathcal{U}(a,b)$ denotes a uniform distribution between $a$ and $b$. We also apply uniform priors on all of our sampled parameters (except the optical depth to reionization $\tau_{\rm reio}$, see Section~\ref{Sec: Data_Sets}). We perform MCMC analysis with \cobaya{}~\cite{COBAYA1,COBAYA2,COBAYA3}, using the Metropolis-Hastings algorithm~\cite{Metropolis_Hastings1,Metropolis_Hastings2}, and consider chains to be converged when the Gelman-Rubin~\cite{Gelman_Rubin} statistic $R-1 \sim 0.05$.
When finding the best fit of the model for a given data combination, we use the \texttt{BOBYQA} algorithm~\cite{BOBYQA1,BOBYQA2} implemented in \cobaya{}. In order to speed up the inference process, we use the emulator \OLE{}~\cite{OLE}\footnote{\url{https://github.com/svenguenther/OLE}}, but we report final results from regular MCMC runs that used \texttt{CLASS} only.

\section{datasets}
\label{Sec: Data_Sets}
For the CMB, we use different combinations of \sptlr{}, \ACTDR{}, and \planck{} data in primary CMB and lensing. We consider constraints from \sptlr{} alone and in combination with \ACTDR{} (\ground{}), and we consider \ground{} in combination with \planck{} (\cmball{}). We also combine each of these data combinations with BAO data from \DESI{}-DR2 (\DESI{}). We list the different datasets involved in this work, following the setup of~\cite{Camphuis_etal}, in Table~\ref{tab:dataset}. 

In all the different combinations, we substitute the low-$\ell\ EE$ information from \planck{} with a prior on $\tau_{\rm reio}$ from the findings of~\cite{planck_collaboration_planck_2020}. This is a Gaussian with a mean value of 0.051 and a standard deviation of 0.006, i.e., $\tau_{\rm reio} \sim \mathcal{N}(0.051, 0.006^2)$. Moreover, we use the CMB-only versions of the likelihoods, known as \texttt{lite}, for the \sptlr{} (\texttt{SPTlite})~\cite{candl} and \ACTDR{} data (\texttt{ACT-lite}), while we use \clipy{}~\cite{Planck_Likelihood} for the \planck{} dataset. When combining \planck{} and \ACTDR{}, we follow the prescription described in~\cite{ACT_DR6_1,ACTDR6_Extended} where the \texttt{lite} version of the \planck{} likelihood, \texttt{plik\_lite}, is used and cut the \planck{} $TT$ spectrum at $\ell>1000$, while $\ell>600$  is cut for the $TE$ and $EE$ spectra. This combination, along with the lensing likelihood of each dataset, is labeled \pact{}-L. The packages associated with each dataset are listed in Table~\ref{tab:dataset_links}.

 We consider \cmball{} as our CMB-only baseline, 
 and \cmball{} + \DESI{} as the one in combination with BAO. We compute $Q_{\rm DMAP}$ and $\Delta$AIC for these two data combinations, and use them to assess the performance of \EDE{}.\footnote{In principle, one can compute \DMAP{} and \AIC{} for \sptlr{} and \ground{}. However, since computing these statistics is computationally expensive, and given that \sptlr{} and \ground{} are subsets of \cmball{} and \cmball{}+\DESI{}, it would add little extra information to assess the model based on \DMAP{} and \AIC{} for such subsets.} When computing $Q_{\rm DMAP}$, we incorporate data from SH0ES as a Gaussian likelihood: $H_0 \sim \mathcal{N}(73.17, 0.86^2)$.
\begin{table}%[hbtp]
    %\resizebox{\textwidth}{!}{%
	\centering
	\begin{tabular}{l|p{0.75\linewidth}}
	\toprule
	\textbf{Name} & \textbf{Dataset} \\
	\midrule

	{\sptlr{}} & \sptlr{} $TT/TE/EE$ CMB spectra~\cite{Camphuis_etal} + CMB lensing spectrum~\cite{MUSE}.\\
	\midrule

	{\planck{}} & Planck 2018 high-$\ell$ $TT/TE/EE$ and low-$\ell$ $TT$ spectra  (PR3)~\cite{Planck_Legacy,Planck2018} + NPIPE PR4 CMB lensing spectrum~\cite{Carron:2022eyg}. \\
	\midrule

	{\ACTDR{}} & \ACTDR{} $TT/TE/EE$ CMB spectra~\cite{ACTDR6_main,ACTDR6_Extended} + \ACTDR{} CMB lensing spectrum~\cite{ACTDR6_Lensing,ACT_Lensing2}.\\

	\midrule
    {\pact{}-L} & The combination of \ACTDR{} and \planck{} with the appropriate $\ell$-cuts for the latter: the $TT$ spectrum cut at $\ell>1000$, while the $TE$ and $EE$ spectra cut at $\ell>600$ + \ACTDR{} CMB lensing spectrum + NPIPE PR4 CMB lensing spectrum.\\
    \midrule
	{\ground{}} & \sptlr{} + \ACTDR{}. \\

	\midrule
	{\cmball{}} & \sptlr{} + \pact{}-L.\\

	\midrule
	{DESI} & DESI-DR2 BAO data~\cite{desi2025}.\\

	\bottomrule
    
	\end{tabular}
   % }
	\caption{Summary of datasets used in this work.}
	\label{tab:dataset}
    
\end{table}

\section{Results}
 \label{Sec:Results}
In this section, we present our main results, starting with constraints from CMB datasets and then in combination with \DESI{}.

\subsection{Constraints from CMB data alone}
\label{Sec:CMB_alone}
\begin{figure*}
\vspace{-1cm}

    \begin{subfigure}%{0.45\textwidth}
        \centering
    \includegraphics[width=0.62\textwidth]{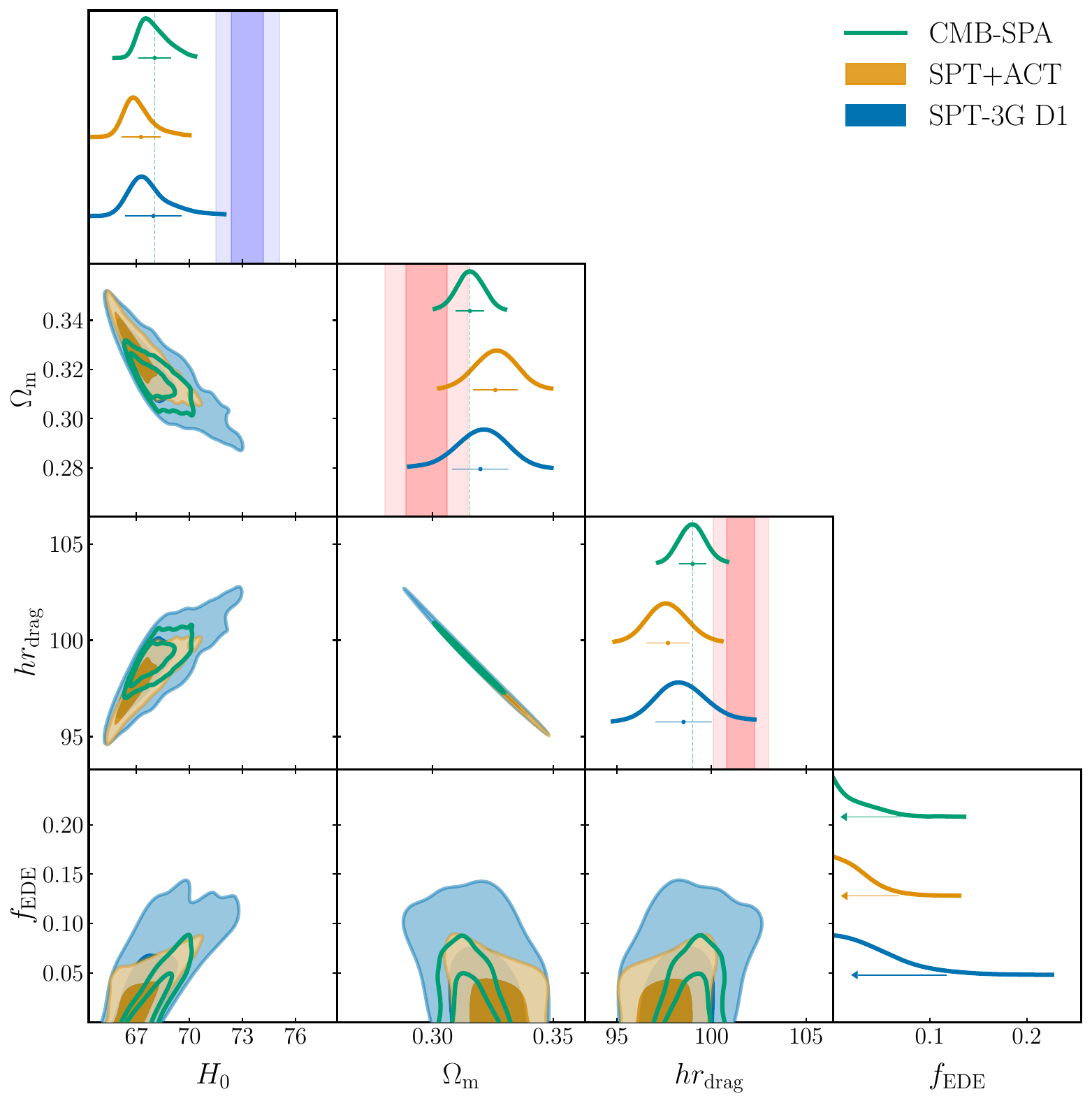}
    \end{subfigure}
    \hfill
    \begin{subfigure}%{0.45\textwidth}
        \centering
    \includegraphics[width=0.62\textwidth]{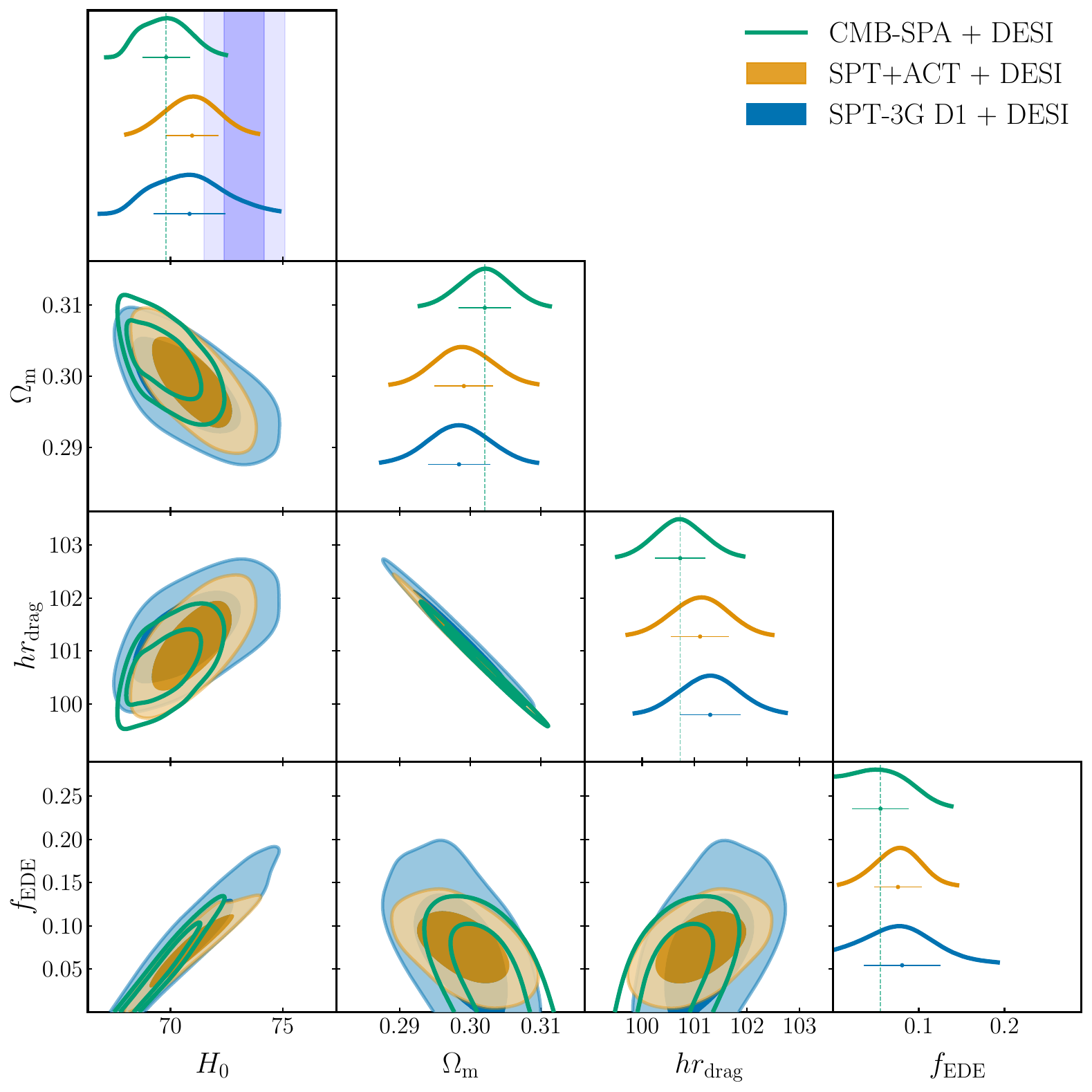}
    \end{subfigure}
    \caption{\textit{Top:} constraints on the parameter set $\{H_0,\Omega_m,hr_d,f_{\rm EDE}\}$ in the \EDE{} model from \sptlr{} in blue, \ground{} in orange, and \cmball{} in green (see Table~\ref{tab:dataset} for the definition of each dataset). The shaded purple band in the $H_0$ 1D posterior corresponds to the SH0ES measurements~\cite{Breuval:2024lsv}, while the red bands appearing in the 1D posteriors of $\Omega_m$ and $hr_d$ correspond to the \DESI{} inferred values within $\Lambda$CDM~\cite{desi2025}. \textit{Bottom}: same as the top plot, but with the inclusion of \DESI{} data. From CMB data, we do not find any preference for EDE over $\Lambda$CDM and the Hubble tension remains, now at the $3.6\, \sigma$ level with \cmball{}. On the other hand, adding \DESI{} data reduces the tension to $2.6\, \sigma$, yet there is still no strong statistical evidence for \EDE{}.}
    \label{fig:Constraints}
\end{figure*}

 We show our constraints from CMB datasets in the top plot of Fig.~\ref{fig:Constraints}\footnote{Although $R-1\,\leq\, 0.05$ for all the MCMC runs appearing in this work, we made a further test of the convergence of these chains. We split the samples into two random sets (after removing the burn-in) and found that the constraints from each set match to an excellent degree.} and the upper part of Table~\ref{Tab:Results}. In addition to $H_0$ and $f_{\rm EDE}$, we present constraints on the matter density parameter, $\Omega_m$, and the sound horizon at the baryonic drag epoch~\cite{BDrag1,Bdrag2}, $r_d$, multiplied by $h=H_0/(100$ km$/{\rm s}/{\rm Mpc})$. 
 %The use of the latter two parameters will become evident in Section~\ref{Sec:CMB_and_BAO}.
 %are particularly interesting to consider given that they are well constrained by BAO data within $\Lambda$CDM and are used to assess the consistency between CMB and BAO data~\cite{Camphuis_etal}.

 From CMB data alone, we first find that different CMB data combinations are consistent with each other in this model space to within $\sim 0.4\, \sigma$. Following~\cite{Camphuis_etal}, we check for the consistency between two datasets by computing $\chi^2\,=\,\Delta p^{\rm T}\Sigma^{-1}\Delta p$ ($\Delta p$ is the difference between the means of a parameter $p$ from two datasets, $\Sigma^{-1}$ is the inverse of the sum of the covariance matrices, and $T$ stands for transpose), from which we compute its PTE and then obtain the corresponding one-dimensional Gaussian fluctuation. This computation is done on the parameters set $\{H_0,\Omega_bh^2,\Omega_ch^2,n_s\}$. Here, $\Omega_b$ and $\Omega_c$ are the density parameters of baryons and dark matter, respectively, and $n_s$ is the spectral index of the primordial power spectrum. 
 
 The upper limit on \fede{} at 95\% confidence level (CL) from \sptlr{} (\fede{}$\, <\,0.12$) is almost twice the corresponding \planck{} one (\fede{}$\, <\, 0.077$), while the \ground{} constraint (\fede{}$\,<0.068$) is 12\% smaller than \planck{}'s.\footnote{We did not compare the constraints from \ground{} to those of the \texttt{CamSpec} likelihood based on \planck{} PR4 maps~\cite{CamSpec}. In that case, we expect \ground{} and \texttt{CamSpec} to give similar constraints.} Furthermore, all datasets appearing in Fig.~\ref{fig:Constraints} are consistent with $f_{\rm EDE}=0$ at less than 68\% CL. With \cmball{}, we get the most up-to-date constraint on \fede{} from the CMB alone, \fede{}$\, <\, 0.070$.\footnote{Note that the slight increase in the upper limit going from \ground{} to \cmball{} is due to including only large-scale information from \planck{}, which tends to shift the posteriors to higher values~\cite{Ups_Downs_EDE,EDE_Silvia_Lennart,EDE_ACT2}.}
 
 To evaluate the impact of adding SPT data on the constraints, specifically those from the P-ACT combination of~\cite{ACTDR6_Extended}, we consider constraints from CMB-SPA without including lensing information from any of the three datasets. The addition of \sptlr{} (without lensing) to the P-ACT dataset reduces the upper limit from \fede{}$\, <\,0.12$~\cite{ACTDR6_Extended} to \fede{}$\, <\,0.091$, a 24\% improvement in constraints. Note that in this comparison we used the \texttt{sroll2} prior on $\tau_{\rm reio}$~\cite{Pagano:2019tci} in accordance with the P-ACT dataset of~\cite{ACTDR6_Extended}.
 %the constraint from \ground{} (\fede{}$\, <\, 0.068$) is around 12\% smaller than the one from \planck{} alone. 

Finally, as can be seen from Table~\ref{Tab:Results}, the Hubble tension gets reduced to $2.3\, \sigma$ and $2.9\, \sigma$ for \sptlr{} and \ground{}, respectively, using $Q_{\rm MPCL}$, compared to $6.3\, \sigma$ and $6.8\, \sigma$ in \LCDM{}, respectively, found in~\cite{Camphuis_etal}. However, with the CMB only baseline dataset, \cmball{}, the tension is still present, albeit reduced from $6.4\, \sigma$ in the $\Lambda$CDM case to $3.6 \, \sigma$ for this model. In addition, we find $Q_{\rm DMAP} = 3.0\, \sigma$ for CMB-SPA, which does not pass our threshold for this test. Moreover, comparing the best-fit $\chi^2$ between this model and $\Lambda$CDM ($\Delta\chi^2=\chi^2_{\rm AEDE} - \chi^2_{\Lambda \rm CDM}$), we find a $\Delta \chi^2 = -5.1$ improvement in the best fit with the \cmball{} dataset. This corresponds to $\Delta{\rm AIC}=0.91$, which does not pass our threshold.  

Therefore, out of the three metrics used here to assess the performance of \EDE{} ($Q_{\rm MPCL},\ Q_{\rm DMAP}$, and $\Delta{\rm AIC}$), we find that it passes none of them. This means that, from CMB data alone, \EDE{} cannot be considered as a solution to the Hubble tension.

% \begin{figure*}%[H]
%     \centering
%     \includegraphics[width=\textwidth]{figures/Tri_Main_SPT3G_D1_SPTACT_CMBSPA.pdf}
%     \caption{Constraints on the set $\{H_0,\Omega_m,hr_d,f_{\rm EDE}\}$ in the \EDE{} model from \sptlr{} in blue, \ground{} in orange and \cmball{} in green (see Table~\ref{tab:dataset} for the definition of each dataset). Below each 1D posterior, we show the mean and 68\% CL of every parameter, except for $f_{\rm EDE}$ for which we show the 95\% CL upper limit. The shaded purple band in the $H_0$ 1D posterior corresponds to the SH0ES measurements~\cite{Breuval:2024lsv}, while the green and red bands appearing in the 1D posteriors of $\Omega_m$ and $hr_d$, respectively, correspond to the \DESI{} inferred values within $\Lambda$CDM~\cite{desi2025}. From CMB data, we do not find any preference for \EDE{} over $\Lambda$CDM and the Hubble tension remains, now at the $3.3\, \sigma$ level with \cmball{}.}
% \end{figure*}

% \begin{figure*}%[H]
%     \centering
%     \includegraphics[width=\textwidth]{figures/Tri_Main_SPT3G_D1_DESI_SPTACT_DESI_CMBSPA_DESI.pdf}
%     \caption{Same as the top plot, but with the inclusion of \DESI{} data. Instead of showing 95\% CL upper limits on $f_{\rm EDE}$, we present its mean and 68\% CL below each 1D posterior. adding \DESI{} data reduces the tension below the $3\, \sigma$ threshold, yet there is still no strong statistical evidence for \EDE{}.}
% \end{figure*}
\begin{table}%[H]
\resizebox{\linewidth}{!}{%
	\begin{tabular}{|c|c|c|c|}
		\hline
		Parameter & SPT-3G D1 & \ground{} & \cmball{} \\
		\hline
$H_0$ [km/s/Mpc] & $67.96^{+0.58}_{-1.7}$ & $67.25^{+0.47}_{-1.1}$ & $68.03^{+0.53}_{-1.1}$ \\
%\hline
$\Omega_{\mathrm{m}}$ & $0.320^{+0.012}_{-0.010}$ & $0.3260\, \pm \, 0.0093$ & $0.3155\,\pm\, 0.0058$ \\
%\hline
$hr_\mathrm{drag}$ [Mpc] & $98.5^{+1.2}_{-1.6}$ & $97.7^{+1.0}_{-1.2}$ & $99.00^{+0.67}_{-0.75}$ \\
%\hline
$f_\mathrm{EDE}$ & $< 0.12$ & $< 0.068$ & $< 0.070$ \\

%\hline
		$Q_\mathrm{MPCL}$[$\sigma$] & $2.3$ & $2.9$ & $3.6$ \\

 %       \hline
        $Q_{\rm DMAP}$[$\sigma$] & $---$ & $---$ & $3.0$ 
        \\
  %      \hline
        $\Delta$AIC & $---$ & $---$ & $0.91$ \\
        \hline
        		\hline

    \multicolumn{3}{c}{\hspace{3cm} + DESI}
    \\
		\hline
        \hline
$H_0$ [km/s/Mpc]& $70.9^{+1.3}_{-1.9}$ & $71.0\, \pm \, 1.2$ & $69.81^{+0.95}_{-1.2}$ \\
%\hline
$\Omega_{\mathrm{m}}$ & $0.2984\, \pm \, 0.0044$ & $0.2991\, \pm \, 0.0042$ & $0.3020\pm 0.0037$ \\
%\hline
$hr_\mathrm{drag}$ [Mpc] & $101.30\, \pm \, 0.58$ & $101.10\, \pm \, 0.55$ & $100.73\pm 0.48$ \\
%\hline

$f_\mathrm{EDE}$ & $0.081^{+0.037}_{-0.052}$ & $0.076^{+0.028}_{-0.027}$ & $0.055^{+0.024}_{-0.047}$ \\
 & $(0.16)$ & $(0.12)$ & $(0.11)$\\

%		\hline
		$Q_\mathrm{MPCL}$[$\sigma$] & $1.5$ & $1.8$ & $2.6$ \\
%		\hline
        $Q_{\rm DMAP}$[$\sigma$] & $---$ & $---$ & $2.4$ 
       \\
 %       \hline
         $\Delta$AIC & $---$ & $---$ & $-2.5$ \\
        \hline
	\end{tabular}
    }
	\caption{\textit{Top}: CMB-only constraints on cosmological parameters for \EDE{}. For $f_{\rm EDE}$, we present upper limits at the 95\% CL. We also show $Q_{\rm MPCL}$ for every dataset, while we show the $Q_{\rm DMAP}$ (eq.~\ref{eq:Q_DMAP}) and $\Delta$AIC (eq.~\ref{eq:AIC}) metrics for \cmball{}. \textit{Bottom:} same as the top part but %with combination with
    with the addition of \DESI{}. For $f_{\rm EDE}$, we show the mean and 68\% CL and below them in parenthesis the one-tail upper limit at 95\% CL.\footnote{We calculate the one-tail upper limit at 95\% CL by integrating the posterior distribution from the boundary $f_{\rm EDE}=0$, in accordance with the 95\% CL quoted in the top part of the Table.}}
    \label{Tab:Results}
\end{table}

\subsection{Constraints from CMB and BAO data}
\label{Sec:CMB_and_BAO}
We now consider the impact of \DESI{} data on \EDE{}, with the results shown in the bottom plot of Fig.~\ref{fig:Constraints} and the lower part of Table~\ref{Tab:Results}. We first assess the consistency between CMB and \DESI{} data by comparing the constraints in the $\Omega_m$-$hr_d$ plane between each CMB dataset mentioned above and \DESI{}.\footnote{We follow the criterion set in~\cite{Camphuis_etal} and consider $3\,\sigma$ as the threshold for two datasets to be consistent (see Section VII.C in~\cite{Camphuis_etal}).} We find the consistency between \sptlr{}, \ground{}, and \cmball{}, with \DESI{} at $1.5\, \sigma$, $2.8\, \sigma$, and $2.5\, \sigma$, respectively, in \EDE{}, compared to $2.3\, \sigma$, $3.7\, \sigma$, and $2.8\, \sigma$ for $\Lambda$CDM~\cite{Camphuis_etal}. Given that all three are below the $3\, \sigma$ threshold, we proceed to combine each CMB dataset with \DESI{}.
 
 When we include \DESI{} data, the conclusions of the previous section change. First, as expected, the degeneracy of $f_{\rm EDE}$ with $H_0$ and $\Omega_ch^2$ (see Figs.~\ref{fig:Additional_CMB_Only} and~\ref{fig:Additional_CMB_BAO}), and the preference for lower (higher) $\Omega_m\ (hr_d)$ by \DESI{}, shifts the constraints with the CMB datasets, rendering them in agreement with the $\Omega_m-hr_d$ constraints (within $\Lambda$CDM) from \DESI{} (see bottom plot of Fig.~\ref{fig:Constraints}). Second, we no longer have an upper limit on $f_{\rm EDE}$, but rather a weak preference for \fede{}$\, >\, 0$, as can be seen from the lower part of Table~\ref{Tab:Results}. These constraints deviate from $f_{\rm EDE}=0$ at $1.6\, \sigma$, $2.8\, \sigma$, and $1.2\, \sigma$
 %are consistent with $f_{\rm EDE}\sim 10\%$ at $0.42\sigma$, $0.86\sigma$ and $0.82\sigma$ 
 for \sptlr{}, \ground{}, and \cmball{} (each combined with \DESI{}), respectively. Note that the increase in the error bar going from \ground{} to \cmball{} seen in Table~\ref{Tab:Results} is attributed to the highly non-Gaussian distributions and the preference for different regions in the $(f_{\rm EDE},\theta_{\rm i})$ plane between the two data combinations (see Fig.~\ref{fig:Additional_CMB_BAO}). Third, the Hubble tension is now reduced below the $3\, \sigma$ threshold for all three CMB datasets when including \DESI{}, with $Q_{\rm MPCL}=2.6\, \sigma$ for \cmball{} + \DESI{}. Furthermore, $Q_{\rm DMAP}=2.4\, \sigma$ for this model with \cmball{} + \DESI{}, and thus passes the threshold for it to be a better fit to the data when SH0ES information is included. Finally, the goodness of fit for this model is also improved compared to $\Lambda$CDM, with $\Delta\chi^2=-8.5$ for \cmball{} + \DESI{}. This improvement is better than other Hubble tension solutions, such as the varying electron mass in a non-flat geometry~\cite{VarMe1,Planck_VarMe,VarMe2015,VarMe2018}, which has $\Delta\chi^2=-6.5$ (see Table VII of~\cite{Camphuis_etal} for comparison with other models). However, the resultant $\Delta$AIC for this model is $\Delta{\rm AIC}=-2.5$, which corresponds to a weak preference for \EDE{} over $\Lambda$CDM and thus does not pass the threshold.

In this paper, we are focused on CMB-only and CMB + BAO analyses, but (uncalibrated) supernova Ia (SNIa) data such as the Dark Energy Survey~\cite{DES:2024jxu}, Pantheon+~\cite{Pantplus1,Pantplus2}, and Union3~\cite{Union} datasets, are relevant as well. They are known to increase the tension with SH0ES in \EDE{} when added to CMB-only data~\cite{Vivian_lates_EDE}. We have therefore, very briefly, explored the implications of adding one of these datasets, Pantheon+, to CMB-SPA + DESI. For this combination, we find $Q_\mathrm{MPCL} = 2.9\, \sigma$. This tension is indeed higher than what we had without Pantheon+, but still passes our $3\, \sigma$ threshold (although just barely).\footnote{A detailed analysis on the impact of SNIa data on the constraints of \EDE{} is beyond the scope of the current work. It is being investigated in a more elaborate analysis in~[Jhaveri et al, in preparation].}

 In summary, the addition of DESI to CMB data does ease the tension according to the $Q_{\rm MPCL}$ and $Q_{\rm DMAP}$ metrics. However, there is no strong statistical evidence to prefer this model over $\Lambda$CDM. Moreover, it is important to point out that this shift in parameters for \EDE{} when including \DESI{} data is another manifestation of the discrepancy that currently exists between CMB and \DESI{} BAO data in \LCDM{}, a discrepancy that could potentially be explained in several ways~\cite{Camphuis_etal}.

\begin{figure}%[H]
    \centering
    \includegraphics[width=\linewidth]{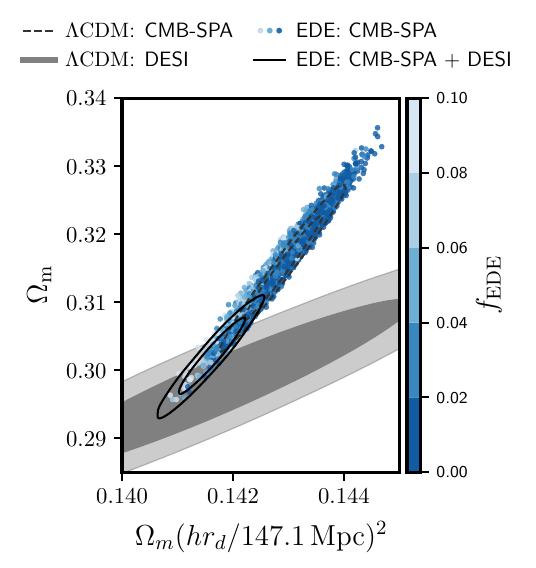}
    \caption{Constraints in the $\left[\Omega_m,\Omega_m(hr_d/147.1 {\rm \ Mpc})^2\right]$ plane from \cmball{} (blue dots) and \cmball{} + DESI (solid black contours) for the \EDE{} model. Also shown are constraints from \cmball{} (gray dashed contours) and \DESI{} (gray solid contours) within \LCDM. (Note that the \DESI{} contours are identical in the \EDE{} case due to the insensitivity of the BAO data to this model's parameters.) The dots are colored according to the $f_{\rm EDE}$ values of the samples in the range $[0,0.1]$. The discrepancy between CMB and \DESI{} BAO data in \LCDM{} projects onto a preference for higher values of $f_{\rm EDE}$.}
    \label{fig:EDE_SPA_DESI}
\end{figure}

 As found in~\cite{Camphuis_etal}, the CMB-BAO discrepancy can be projected onto many extensions of $\Lambda$CDM, and \EDE{} is yet another example. To see this more clearly, we plot in Fig.~\ref{fig:EDE_SPA_DESI} the 2D posterior of $\left[\Omega_m,\Omega_m(hr_d/147.1 \ {\rm Mpc})^2\right]$ for \cmball{} and \cmball{} + \DESI{}, within \EDE{}, and the posterior for \DESI{} alone and \cmball{} within $\Lambda$CDM. The reason for choosing $\Omega_m(hr_d/147.1\ {\rm Mpc})^2$ instead of $hr_d$ is twofold. First, it allows for a better visualization of the discrepancy, given the strong $\Omega_m-hr_d$ degeneracy, and second $\Omega_m(hr_d)^2$ is a combination 
%that both the CMB and BAO data are sensitive to.
which both the CMB and BAO data constrain well. From this plot, we can see how allowing $f_{\rm EDE}$ to take on non-zero values (roughly $4-10\%$) results in an overlap between the \cmball- and \DESI-preferred regions, thus reducing the discrepancy seen in $\Lambda$CDM. This explains why adding \DESI{} data results in the shifts seen in Table~\ref{Tab:Results} and an improvement in the fit compared to $\Lambda$CDM.
% \begin{table}%[H]
% 	\begin{tabular}{|c|c|c|c|}
% 		\hline
% 		Parameter & SPT-3G D1 & \ground{} & \cmball{} \\
% 		\hline
% 		$H_0$ & $70.9^{+1.3}_{-1.9}$ & $71.0\pm 1.2$ & $70.3\pm 1.1$ \\
% 		\hline
% 		$\Omega_{\mathrm{m}}$ & $0.2984\pm 0.0044$ & $0.2991\pm 0.0042$ & $0.3012\pm 0.0037$ \\
% 		\hline
% 		$hr_\mathrm{drag}$ & $101.30\pm 0.58$ & $101.10\pm 0.55$ & $100.82\pm 0.49$ \\
% 		\hline
% 		$f_{\rm EDE}$ & $0.081^{+0.037}_{-0.052}$ & $0.076\pm 0.028$ & $0.072\pm 0.034$ \\
% 		\hline
% 		$Q_\mathrm{MPCL}$ & $1.5\, \sigma$ & $1.8\, \sigma$ & $2.3\, \sigma$ \\
% 		\hline
% 	\end{tabular}
%     \label{Tab:Results}
% 	\caption{CMB and DESI}
% \end{table}

 \section{Conclusion}
 \label{Sec:Conclusion}

In this work, we extend the cosmological results of~\cite{Camphuis_etal} to include constraints on the \EDE{} model. We focus on constraints from \sptlr{}, \ground{}, and \cmball{}. We also present constraints from adding \DESI{} to each of these data combinations.

From CMB data alone, we find that all data combinations are statistically consistent with each other and do not show appreciable evidence for \EDE{}. The constraints on \fede{} coming from \ground{} are $\sim 12\%$ smaller than the ones from \planck{}. Moreover, with \cmball{}, we find that the Hubble tension is still present at the $3.6\, \sigma$ level ($H_0 = 68.03^{+0.53}_{-1.1}$) with the upper limit on $f_{\rm EDE}$ at the 95\% CL being $f_{\rm EDE} < 0.070$, and the improvement in best fit for this case compared to \LCDM{} is only 5.1 $\chi^2$ points. With the $Q_{\rm DMAP}$ metric for this data combination also not passing our threshold ($Q_{\rm DMAP} = 3.0 \, \sigma$), these results show that, from CMB data alone, \EDE{} is not able to solve the tension and does not show a strong statistical preference over \LCDM{}.

We also find that the discrepancy between \sptlr{}, \ground{}, and \cmball with \DESI{} in the $\Omega_m$-$hr_d$ plane is reduced to $1.5\, \sigma$, $2.8\, \sigma$, and $2.5\, \sigma$, respectively, for this model. With this level of agreement within the \EDE{} model, the two types of data can be sensibly combined.

\begin{figure*}

\centering
%\hspace{-5cm}
    \includegraphics[width=1.\textwidth]{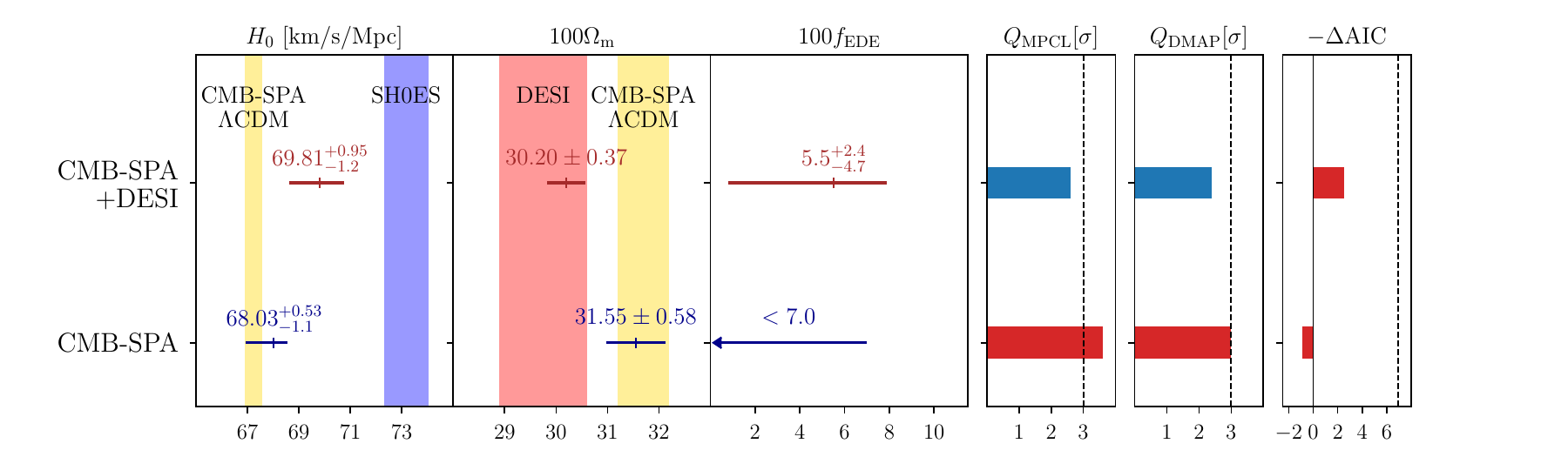}
    \caption{Summary of the main results of this work. Shown are constraints, within \EDE{}, from \cmball{} (in dark blue) and \cmball{} + \DESI{} (in brown) on $H_0$ (far left), $100\Omega_m$ (second to the left), and \fede{} (third to the left). In the $H_0$ panel, we show the 68\% CL from SH0ES~\cite{Breuval:2024lsv} (purple band) and that of \cmball{} within \LCDM{}~\cite{Camphuis_etal} (gold band). In the $100\Omega_m$ panel, we show the 68\% CL from DESI~\cite{desi2025} (red band) and that of \cmball{} (gold band), both within \LCDM{}. We also show the three metrics \MPCL{} (third to the right), \DMAP{} (second to the right), and $-$\AIC{} (far right). For each of these, we show their passing threshold as dashed lines ($3\, \sigma$, $3\, \sigma$, and $6.91$, respectively), and in blue the cases that pass these thresholds while in red those that do not. More details are presented in Table~\ref{Tab:Results}.}
    \label{fig:money_plot}
\end{figure*}

Unlike the CMB-only case, adding BAO data from \DESI{} results in a mild preference for $f_{\rm EDE}\, >\,0$ which, for \cmball{} + \DESI{} is $f_{\rm EDE} = 0.055^{+0.024}_{-0.047}$; a deviation from $f_{\rm EDE} = 0$ at $1.2\, \sigma$. This preference is accompanied with an increase in $H_0$ to $69.81^{+0.95}_{-1.2}$ km/s/Mpc, which corresponds to a $2.6\, \sigma$ difference with the SH0ES measurement. With this data combination, the model also passes the $Q_{\rm DMAP}<3\, \sigma$ test, where we find $Q_{\rm DMAP}=2.4\, \sigma$. Moreover, with \cmball{} + \DESI{}, the best-fit $\chi^2$ of this model improves by 8.5 points compared to \LCDM{}. However, this is not enough improvement for the model to pass the $\Delta$AIC criterion ($\Delta{\rm AIC}=-2.5$ for \cmball{} + \DESI{}); at best, the 
%model shows a weak preference  
data indicate a weak preference for the \EDE{} model compared to \LCDM{}. We show a summary of the main results in Fig.~\ref{fig:money_plot}.

%This feature of projecting the discrepancy between CMB and BAO onto extended models has to be interpreted with caution. Of course it could still point to new physics, but one needs strong statistical evidence for that before making any conclusions.
We point out that this shift in parameters seen when adding the \DESI{} data is a manifestation of the currently existing discrepancy between \DESI{} and CMB data in \LCDM{}. With more precise data, such as that expected soon from further SPT releases \cite{Prabhu_etal}, early observations with the Simons Observatory~\cite{SimonsObservatory:2025wwn}, or upcoming data from \DESI{} we might be able to understand the origin of this discrepancy and whether it points to truly new physics or to unknown sources of systematic error.
\section{Acknowledgments}
\label{sec:acknowledgements}
% main SPT
We thank Antony Lewis, Tristan Smith and Vivian Poulin for fruitful comments. The South Pole Telescope program is supported by the National Science Foundation (NSF) through awards OPP-1852617 and OPP-2332483. Partial support is also provided by the Kavli Institute of Cosmological Physics at the University of Chicago. 
% Argonne
Argonne National Laboratory’s work was supported by the U.S. Department of Energy, Office of High Energy Physics, under contract DE-AC02-06CH11357. 
% Davis
The UC Davis group acknowledges support from Michael and Ester Vaida. 
% Fermilab
Work at the Fermi National Accelerator Laboratory (Fermilab), a U.S. Department of Energy, Office of Science, Office of High Energy Physics HEP User Facility, is managed by Fermi Forward Discovery Group, LLC, acting under Contract No. 89243024CSC000002.
% Melbourne
The Melbourne authors acknowledge support from the Australian Research Council’s Discovery Project scheme (No. DP210102386). 
% Paris
The Paris group has received funding from the European Research Council (ERC) under the European Union’s Horizon 2020 research and innovation program (grant agreement No 101001897), and funding from the Centre National d’Etudes Spatiales. 
% SLAC
The SLAC group is supported in part by the Department of Energy at SLAC National Accelerator Laboratory, under contract DE-AC02-76SF00515.

This work has made use of the Infinity Cluster hosted by Institut d'Astrophysique de Paris. We thank Stephane Rouberol for smoothly running this cluster for us. This work relied on the \texttt{NumPy} library for numerical computations~\citep{numpy}, the \texttt{SciPy} library for scientific computing~\citep{scipy}, and the \texttt{Matplotlib} library for plotting~\citep{matplotlib}. Posterior sampling analysis and plotting were performed using the \texttt{GetDist} package~\cite{GetDist}.

\appendix
\section{Additional Results}
\label{Sec:Appen_Add_Res}
\begin{figure*}
    \centering
    \hspace{-1.9cm}
    \includegraphics[width=1.1\textwidth]{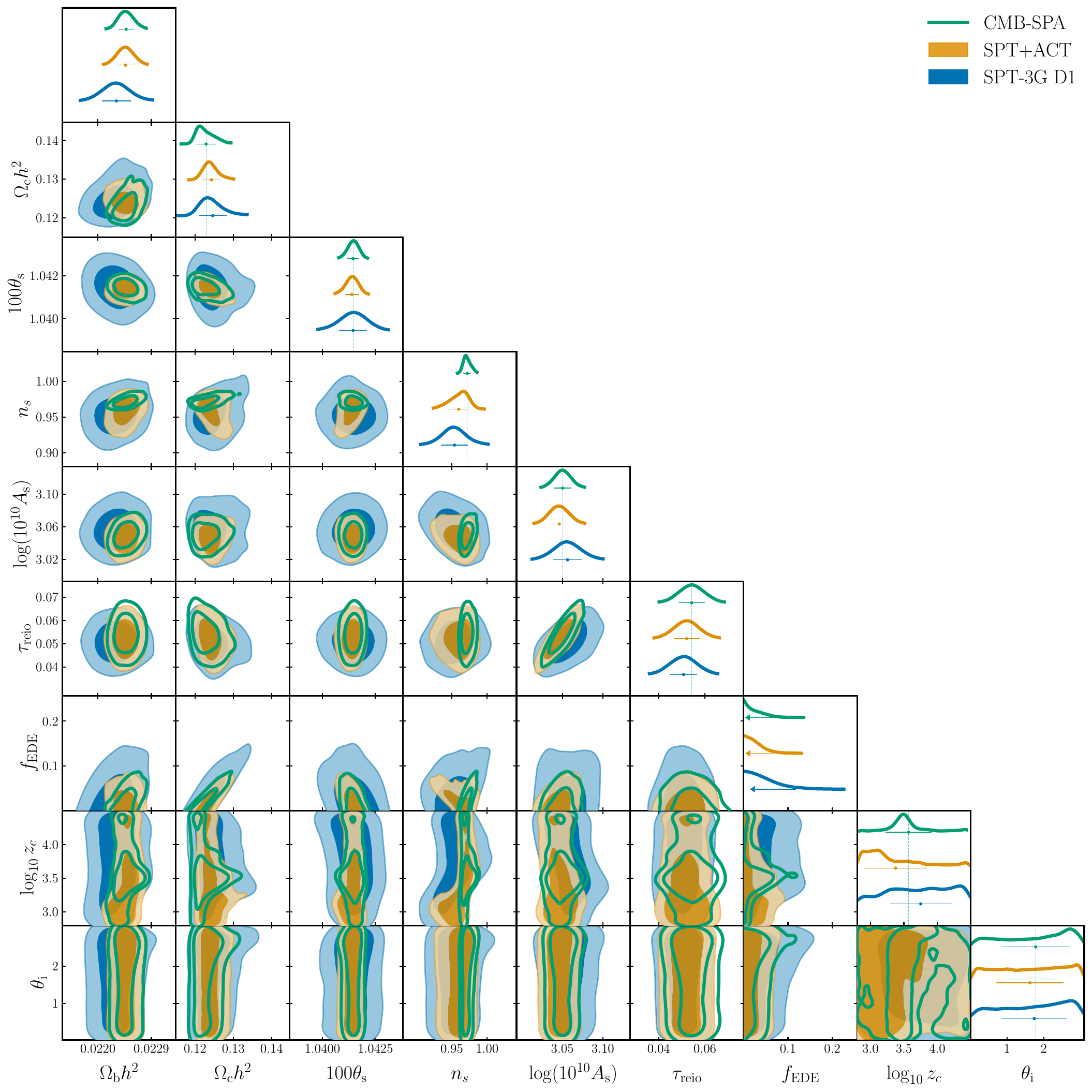}
    \caption{Same as the top plot of Fig.~\ref{fig:Constraints} but for additional cosmological parameters appearing in Table~\ref{Tab:Additional_Parmas}.}
    \label{fig:Additional_CMB_Only}
\end{figure*}

\begin{figure*}
    \centering
    \hspace{-1.9cm}
    \includegraphics[width=1.1\textwidth]{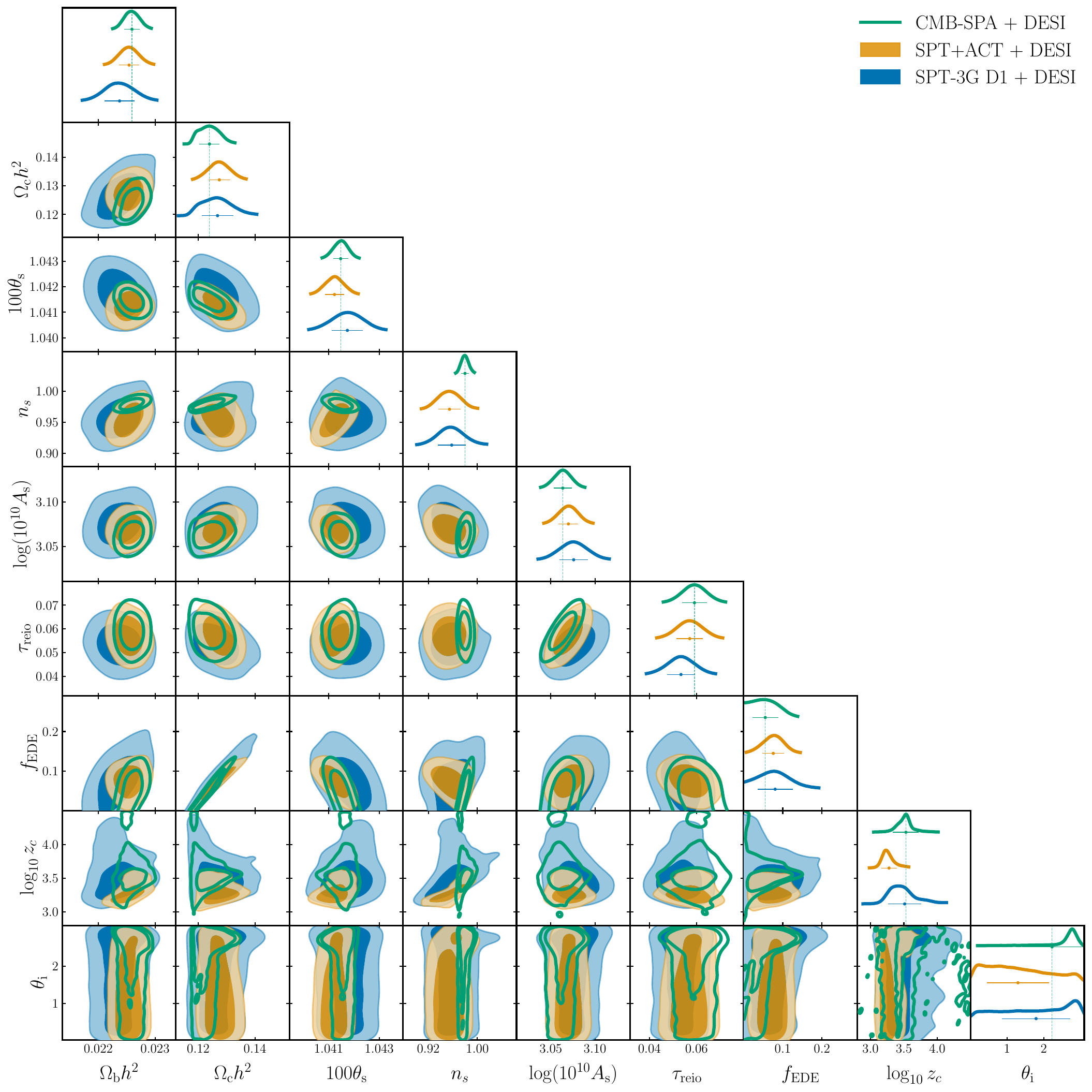}
    \caption{Same as Fig.~\ref{fig:Additional_CMB_Only} but when including \DESI{} data.}
    \label{fig:Additional_CMB_BAO}
\end{figure*}

In this section, for completeness, we present constraints on the set of cosmological parameters $\left\{\Omega_bh^2,\Omega_ch^2,100\theta_s, n_s, \log(10^{10}A_s),f_{\rm EDE},\log_{10}z_c,\theta_{\rm i}\right\}$, where $\theta_s$ is the angular size of the sound horizon at recombination and $A_s$ is the amplitude of the primordial spectrum. The constraints from CMB data only are shown in Fig.~\ref{fig:Additional_CMB_Only}, and those from the combination of CMB and \DESI{} BAO data are shown in Fig.~\ref{fig:Additional_CMB_BAO}. We also present the numerical values of these constraints in Table~\ref{Tab:Additional_Parmas}.

\begin{table*}%[H]
\begin{tabular}{|c|c|c|c|}
\hline
Parameter & SPT-3G D1 & SPT+ACT & CMB-SPA \\
\hline
$\Omega_\mathrm{b}h^2$ & $0.02231\pm 0.00024$ & $0.02246\pm 0.00015$ & $0.02248^{+0.00012}_{-0.00014}$ \\
\hline
$\Omega_\mathrm{c}h^2$ & $0.1246^{+0.0020}_{-0.0039}$ & $0.1243^{+0.0014}_{-0.0025}$ & $0.1229^{+0.0015}_{-0.0031}$ \\
\hline
$100\theta_\mathrm{s}$ & $1.04144^{+0.00069}_{-0.00062}$ & $1.04138^{+0.00033}_{-0.00028}$ & $1.04145^{+0.00029}_{-0.00025}$ \\
\hline
$n_s$ & $0.954^{+0.017}_{-0.019}$ & $0.960^{+0.016}_{-0.011}$ & $0.9718^{+0.0042}_{-0.0065}$ \\
\hline
$\log(10^{10} A_\mathrm{s})$ & $3.056\pm 0.018$ & $3.046\pm 0.013$ & $3.050\pm 0.011$ \\
\hline
$\tau_\mathrm{reio}$ & $0.0508\pm 0.0059$ & $0.0521\pm 0.0057$ & $0.0543\pm 0.0057$ \\
\hline
$f_\mathrm{EDE}$ & $< 0.12$ & $< 0.068$ & $< 0.07$ \\
\hline
$\log_{10}z_c$ & $3.75^{+0.68}_{-0.48}$ & $3.38^{+0.21}_{-0.57}$ & $3.57^{+0.17}_{-0.33}$ \\
\hline
$\theta_\mathrm{i}$ & $1.74^{+1.3}_{-0.83}$ & $1.62\pm 0.91$ & $1.8^{+1.2}_{-1.6}$ \\
        		\hline
\hline
    \multicolumn{3}{c}{\hspace{3cm} + DESI}
    \\
		\hline
        \hline
% \hline
% \end{tabular}
% \caption{Cosmo Params}
% \end{table}
% \begin{table*}%[H]
% \begin{tabular}{|c|c|c|c|}
% \hline
% Parameter & SPT-3G D1 + DESI & SPT+ACT + DESI & CMB-SPA + DESI \\
% \hline
$\Omega_\mathrm{b}h^2$ & $0.02237\pm 0.00026$ & $0.02254\pm 0.00017$ & $0.02259\pm 0.00014$ \\
\hline
$\Omega_\mathrm{c}h^2$ & $0.1268^{+0.0044}_{-0.0066}$ & $0.1274\pm 0.0038$ & $0.1240^{+0.0031}_{-0.0043}$ \\
\hline
$100\theta_\mathrm{s}$ & $1.04174\pm 0.00061$ & $1.04123\pm 0.00038$ & $1.04148\pm 0.00029$ \\
\hline
$n_s$ & $0.958^{+0.020}_{-0.023}$ & $0.955\pm 0.018$ & $0.9798^{+0.0055}_{-0.0066}$ \\
\hline
$\log(10^{10} A_\mathrm{s})$ & $3.076\pm 0.016$ & $3.070\pm 0.011$ & $3.064\pm 0.010$ \\
\hline
$\tau_\mathrm{reio}$ & $0.0534\pm 0.0059$ & $0.0571\pm 0.0056$ & $0.0591\pm 0.0053$ \\
\hline
$f_\mathrm{EDE}$ & $0.081^{+0.037}_{-0.052}$ & $0.076^{+0.028}_{-0.027}$ & $0.055^{+0.024}_{-0.047}$ \\
\hline
$\log_{10}z_c$ & $3.51^{+0.11}_{-0.28}$ & $3.278^{+0.058}_{-0.12}$ & $3.530^{+0.076}_{-0.14}$ \\
\hline
$\theta_\mathrm{i}$ & $1.79^{+1.3}_{-0.63}$ & $1.30^{+0.48}_{-1.2}$ & $2.22^{+0.83}_{-0.12}$ \\
\hline
\end{tabular}
\caption{\textit{Top:} CMB-only constraints on all the parameters of \EDE{} (6 \LCDM{} ones and the 3 specific to \EDE{}. The mean and 68 \% CL range is shown for all parameters except $f_\mathrm{EDE}$, for which we show the upper limit at 95\% CL. \textit{Bottom:} same as the top part but with the addition of DESI. For $f_\mathrm{EDE}$, we show the mean and 68\% CL.}
\label{Tab:Additional_Parmas}
\end{table*}

\section{Packages of datasets}
\label{Sec:Appen_Data}
\begin{nolinenumbers}
\begin{table*}%[hbtp]
    \resizebox{\textwidth}{!}{%
	\centering
	\begin{tabular}{l|p{0.7\textwidth}}
	\toprule
	\textbf{Dataset} & \textbf{Link to package} \\

	\midrule

	{\sptlr{}} & \url{https://pole.uchicago.edu/public/data/camphuis25/} \newline \url{https://github.com/SouthPoleTelescope/spt_candl_data}\newline
    \url{https://pole.uchicago.edu/public/data/ge25/}
    \\
	\midrule

	{\planck{}} & \url{https://github.com/benabed/clipy} \newline
    \url{https://pla.esac.esa.int/}
    \newline\url{https://github.com/carronj/planck_PR4_lensing}
    \\
	\midrule

	{\ACTDR{}} & \url{https://github.com/ACTCollaboration/DR6-ACT-lite} \newline
    \url{https://github.com/ACTCollaboration/act_dr6_lenslike}
    \\

	\bottomrule
    
	\end{tabular}
    }
	\caption{Likelihoods and their corresponding links.}
	\label{tab:dataset_links}

\end{table*}
\end{nolinenumbers}
\bibliography{biblio}

@book{Bayesian1,
        author = "Andreon, Stefano and Weaver, Brian",
        title = "{Bayesian Methods for the Physical Sciences Learning from Examples in Astronomy and Physics}",
        isbn = "978-3-319-15286-8",
        publisher = "Springer Series in Astrostatistics",
        month = "May",
        year = "2015"

}

@book{Jeffreys1939-JEFTOP-5,
	address = {Oxford, England},
	author = {Harold Jeffreys},
	editor = {},
	title = {Theory of Probability},
	year = {1939}
}

@article{Verde1,
    author = "Verde, Licia and Protopapas, Pavlos and Jimenez, Raul",
    title = "{Planck and the local Universe: Quantifying the tension}",
    eprint = "1306.6766",
    archivePrefix = "arXiv",
    primaryClass = "astro-ph.CO",
    doi = "10.1016/j.dark.2013.09.002",
    journal = "Phys. Dark Univ.",
    volume = "2",
    pages = "166--175",
    year = "2013"
}

@article{In_Realm,
    author = "Di Valentino, Eleonora and Mena, Olga and Pan, Supriya and Visinelli, Luca and Yang, Weiqiang and Melchiorri, Alessandro and Mota, David F. and Riess, Adam G. and Silk, Joseph",
    title = "{In the realm of the Hubble tension\textemdash{}a review of solutions}",
    eprint = "2103.01183",
    archivePrefix = "arXiv",
    primaryClass = "astro-ph.CO",
    reportNumber = "IPPP/20/108",
    doi = "10.1088/1361-6382/ac086d",
    journal = "Class. Quant. Grav.",
    volume = "38",
    number = "15",
    pages = "153001",
    year = "2021"
}

@article{H0_Olympics,
    author = {Sch\"oneberg, Nils and Franco Abell\'an, Guillermo and P\'erez S\'anchez, Andrea and Witte, Samuel J. and Poulin, Vivian and Lesgourgues, Julien},
    title = "{The H0 Olympics: A fair ranking of proposed models}",
    eprint = "2107.10291",
    archivePrefix = "arXiv",
    primaryClass = "astro-ph.CO",
    doi = "10.1016/j.physrep.2022.07.001",
    journal = "Phys. Rept.",
    volume = "984",
    pages = "1--55",
    year = "2022"
}

@ARTICLE{VerdeTreuRiess,
	author = {{Verde}, Licia and {Treu}, Tommaso and {Riess}, Adam G.},
	title = "{Tensions between the early and late Universe}",
	journal = {Nature Astronomy},
	year = "2019",
	month = "Sep",
	volume = {3},
	pages = {891-895},
	doi = {10.1038/s41550-019-0902-0},
	archivePrefix = {arXiv},
	eprint = {1907.10625},
	primaryClass = {astro-ph.CO},
	adsurl = {https://ui.adsabs.harvard.edu/abs/2019NatAs...3..891V}
}

@article{Camarena_2019,
    author = "Camarena, David and Marra, Valerio",
    title = "{Local determination of the Hubble constant and the deceleration parameter}",
    eprint = "1906.11814",
    archivePrefix = "arXiv",
    primaryClass = "astro-ph.CO",
    doi = "10.1103/PhysRevResearch.2.013028",
    journal = "Phys. Rev. Res.",
    volume = "2",
    number = "1",
    pages = "013028",
    year = "2020"
}

@article{Camarena_2021,
    author = "Camarena, David and Marra, Valerio",
    title = "{On the use of the local prior on the absolute magnitude of Type Ia supernovae in cosmological inference}",
    eprint = "2101.08641",
    archivePrefix = "arXiv",
    primaryClass = "astro-ph.CO",
    doi = "10.1093/mnras/stab1200",
    journal = "Mon. Not. Roy. Astron. Soc.",
    volume = "504",
    number = "4",
    pages = "5164--5171",
    year = "2021"
}

@article{Planck2018,
	author = "Aghanim, N. and others",
	collaboration = "Planck",
	title = "{Planck 2018 results. VI. Cosmological parameters}",
	eprint = "1807.06209",
	archivePrefix = "arXiv",
	primaryClass = "astro-ph.CO",
	doi = "10.1051/0004-6361/201833910",
	journal = "Astron. Astrophys.",
	volume = "641",
	pages = "A6",
	year = "2020"
}

@article{Efstathiou_Poulin_EDE,
    author = "Efstathiou, George and Rosenberg, Erik and Poulin, Vivian",
    title = "{Improved Planck Constraints on Axionlike Early Dark Energy as a Resolution of the Hubble Tension}",
    eprint = "2311.00524",
    archivePrefix = "arXiv",
    primaryClass = "astro-ph.CO",
    doi = "10.1103/PhysRevLett.132.221002",
    journal = "Phys. Rev. Lett.",
    volume = "132",
    number = "22",
    pages = "221002",
    year = "2024"
}

@INPROCEEDINGS{DESI,
	collaboration = "{DESI collaboration}",
	title = "{The Dark Energy Spectroscopic Instrument (DESI)}",
	booktitle = {Bulletin of the American Astronomical Society},
	year = 2019,
	volume = {51},
	month = sep,
	eid = {57},
	pages = {57},
	archivePrefix = {arXiv},
	eprint = {1907.10688},
	primaryClass = {astro-ph.IM},
	adsurl = {https://ui.adsabs.harvard.edu/abs/2019BAAS...51g..57L}
}

@article{DES:2024jxu,
    author = "Abbott, T. M. C. and others",
    collaboration = "DES",
    title = "{The Dark Energy Survey: Cosmology Results with \ensuremath{\sim}1500 New High-redshift Type Ia Supernovae Using the Full 5 yr Data Set}",
    eprint = "2401.02929",
    archivePrefix = "arXiv",
    primaryClass = "astro-ph.CO",
    reportNumber = "FERMILAB-PUB-23-0821-PPD, DES-2023-805",
    doi = "10.3847/2041-8213/ad6f9f",
    journal = "Astrophys. J. Lett.",
    volume = "973",
    number = "1",
    pages = "L14",
    year = "2024"
}

@article{Union,
    author = "Rubin, David and others",
    title = "{Union Through UNITY: Cosmology with 2,000 SNe Using a Unified Bayesian Framework}",
    eprint = "2311.12098",
    archivePrefix = "arXiv",
    primaryClass = "astro-ph.CO",
    doi = "10.3847/1538-4357/adc0a5",
    journal = "Astrophys. J.",
    volume = "986",
    number = "2",
    pages = "231",
    year = "2025"
}

@article{Trouble_H0,
    author = "Bernal, Jose Luis and Verde, Licia and Riess, Adam G.",
    title = "{The trouble with $H_0$}",
    eprint = "1607.05617",
    archivePrefix = "arXiv",
    primaryClass = "astro-ph.CO",
    doi = "10.1088/1475-7516/2016/10/019",
    journal = "JCAP",
    volume = "10",
    pages = "019",
    year = "2016"
}

@article{Planck_Likelihood,
    author = "Aghanim, N. and others",
    collaboration = "Planck",
    title = "{Planck 2018 results. V. CMB power spectra and likelihoods}",
    eprint = "1907.12875",
    archivePrefix = "arXiv",
    primaryClass = "astro-ph.CO",
    doi = "10.1051/0004-6361/201936386",
    journal = "Astron. Astrophys.",
    volume = "641",
    pages = "A5",
    year = "2020"
}

@article{ACT_DR6_lens,
    author = "Madhavacheril, Mathew S. and others",
    collaboration = "ACT",
    title = "{The Atacama Cosmology Telescope: DR6 Gravitational Lensing Map and Cosmological Parameters}",
    eprint = "2304.05203",
    archivePrefix = "arXiv",
    primaryClass = "astro-ph.CO",
    reportNumber = "FERMILAB-PUB-23-206-PPD",
    month = "4",
    year = "2023"
}

@article{EDE1,
    author = "Poulin, Vivian and Smith, Tristan L. and Karwal, Tanvi and Kamionkowski, Marc",
    title = "{Early Dark Energy Can Resolve The Hubble Tension}",
    eprint = "1811.04083",
    archivePrefix = "arXiv",
    primaryClass = "astro-ph.CO",
    doi = "10.1103/PhysRevLett.122.221301",
    journal = "Phys. Rev. Lett.",
    volume = "122",
    number = "22",
    pages = "221301",
    year = "2019"
}

@article{EDE2,
    author = {M\"ortsell, Edvard and Dhawan, Suhail},
    title = "{Does the Hubble constant tension call for new physics?}",
    eprint = "1801.07260",
    archivePrefix = "arXiv",
    primaryClass = "astro-ph.CO",
    doi = "10.1088/1475-7516/2018/09/025",
    journal = "JCAP",
    volume = "09",
    pages = "025",
    year = "2018"
}

@article{EDE3,
    author = "Karwal, Tanvi and Kamionkowski, Marc",
    title = "{Dark energy at early times, the Hubble parameter, and the string axiverse}",
    eprint = "1608.01309",
    archivePrefix = "arXiv",
    primaryClass = "astro-ph.CO",
    doi = "10.1103/PhysRevD.94.103523",
    journal = "Phys. Rev. D",
    volume = "94",
    number = "10",
    pages = "103523",
    year = "2016"
}

@article{CLASS1,
    author = "Lesgourgues, Julien",
    title = "{The Cosmic Linear Anisotropy Solving System (CLASS) I: Overview}",
    eprint = "1104.2932",
    archivePrefix = "arXiv",
    primaryClass = "astro-ph.IM",
    month = "4",
    year = "2011"
}

@article{CLASS2,
    author = "Blas, Diego and Lesgourgues, Julien and Tram, Thomas",
    title = "{The Cosmic Linear Anisotropy Solving System (CLASS) II: Approximation schemes}",
    eprint = "1104.2933",
    archivePrefix = "arXiv",
    primaryClass = "astro-ph.CO",
    reportNumber = "CERN-PH-TH-2011-082, LAPTH-010-11",
    doi = "10.1088/1475-7516/2011/07/034",
    journal = "JCAP",
    volume = "07",
    pages = "034",
    year = "2011"
}

@article{AxiCLASS1, 
        author = "Smith, Tristan L. and Poulin, Vivian and Amin, Mustafa A.", 
        title = "{Oscillating scalar fields and the Hubble tension: a resolution with novel signatures}", 
        eprint = "1908.06995", 
        archivePrefix = "arXiv", 
        primaryClass = "astro-ph.CO", 
        doi = "10.1103/PhysRevD.101.063523", 
        journal = "Phys. Rev. D", 
        volume = "101", 
        number = "6", 
        pages = "063523", 
        year = "2020" 
}

@article{Pagano:2019tci,
    author = "Pagano, L. and Delouis, J. -M. and Mottet, S. and Puget, J. -L. and Vibert, L.",
    title = "{Reionization optical depth determination from Planck HFI data with ten percent accuracy}",
    eprint = "1908.09856",
    archivePrefix = "arXiv",
    primaryClass = "astro-ph.CO",
    doi = "10.1051/0004-6361/201936630",
    journal = "Astron. Astrophys.",
    volume = "635",
    pages = "A99",
    year = "2020"
}

@article{AxiCLASS2, author = "Poulin, Vivian and Smith, Tristan L. and Grin, Daniel and Karwal, Tanvi and Kamionkowski, Marc", title = "{Cosmological implications of ultralight axionlike fields}", eprint = "1806.10608", archivePrefix = "arXiv", primaryClass = "astro-ph.CO", doi = "10.1103/PhysRevD.98.083525", journal = "Phys. Rev. D", volume = "98", number = "8", pages = "083525", year = "2018" 
}

@article{COBAYA1,
    author = "Torrado, Jesus and Lewis, Antony",
    title = "{Cobaya: Code for Bayesian Analysis of hierarchical physical models}",
    eprint = "2005.05290",
    archivePrefix = "arXiv",
    primaryClass = "astro-ph.IM",
    reportNumber = "TTK-20-15",
    doi = "10.1088/1475-7516/2021/05/057",
    journal = "JCAP",
    volume = "05",
    pages = "057",
    year = "2021"
}

@ARTICLE{COBAYA2,
       author = {{Torrado}, Jes{\'u}s and {Lewis}, Antony},
        title = "{Cobaya: code for Bayesian analysis of hierarchical physical models}",
      journal = {JCAP},
         year = 2021,
        month = may,
       volume = {2021},
       number = {5},
          eid = {057},
        pages = {057},
          doi = {10.1088/1475-7516/2021/05/057},
archivePrefix = {arXiv},
       eprint = {2005.05290},
 primaryClass = {astro-ph.IM},
       adsurl = {https://ui.adsabs.harvard.edu/abs/2021JCAP...05..057T}
}

@MISC{COBAYA3,
       author = {{Torrado}, Jes{\'u}s and {Lewis}, Antony},
        title = "{Cobaya: Bayesian analysis in cosmology}",
 howpublished = {Astrophysics Source Code Library, record ascl:1910.019},
         year = 2019,
        month = oct,
          eid = {ascl:1910.019},
        pages = {ascl:1910.019},
archivePrefix = {ascl},
       eprint = {1910.019},
       adsurl = {https://ui.adsabs.harvard.edu/abs/2019ascl.soft10019T}
}

@article{GetDist,
 author         = "Lewis, Antony",
 title          = "{GetDist: a Python package for analysing Monte Carlo
                   samples}",
 year           = "2019",
 eprint         = "1910.13970",
 archivePrefix  = "arXiv",
 primaryClass   = "astro-ph.IM",
 SLACcitation   = "%%CITATION = ARXIV:1910.13970;%%",
 url            = "https://getdist.readthedocs.io"
}

@article{AxiString,
    author = "Kamionkowski, Marc and Pradler, Josef and Walker, Devin G. E.",
    title = "{Dark energy from the string axiverse}",
    eprint = "1409.0549",
    archivePrefix = "arXiv",
    primaryClass = "hep-ph",
    reportNumber = "SLAC-PUB-16085",
    doi = "10.1103/PhysRevLett.113.251302",
    journal = "Phys. Rev. Lett.",
    volume = "113",
    number = "25",
    pages = "251302",
    year = "2014"
}

@article{NEDE,
    author = "Niedermann, Florian and Sloth, Martin S.",
    title = "{Resolving the Hubble tension with new early dark energy}",
    eprint = "2006.06686",
    archivePrefix = "arXiv",
    primaryClass = "astro-ph.CO",
    doi = "10.1103/PhysRevD.102.063527",
    journal = "Phys. Rev. D",
    volume = "102",
    number = "6",
    pages = "063527",
    year = "2020"
}

@article{Elisa-Eiishiro,
    author = "Herold, Laura and Ferreira, Elisa G. M. and Komatsu, Eiichiro",
    title = "{New Constraint on Early Dark Energy from Planck and BOSS Data Using the Profile Likelihood}",
    eprint = "2112.12140",
    archivePrefix = "arXiv",
    primaryClass = "astro-ph.CO",
    doi = "10.3847/2041-8213/ac63a3",
    journal = "Astrophys. J. Lett.",
    volume = "929",
    number = "1",
    pages = "L16",
    year = "2022"
}

@article{Elisa-Laura,
    author = "Herold, Laura and Ferreira, Elisa G. M.",
    title = "{Resolving the Hubble tension with Early Dark Energy}",
    eprint = "2210.16296",
    archivePrefix = "arXiv",
    primaryClass = "astro-ph.CO",
    month = "10",
    year = "2022"
}

@article{Birefrengence1,
    author = "Murai, Kai and Naokawa, Fumihiro and Namikawa, Toshiya and Komatsu, Eiichiro",
    title = "{Isotropic cosmic birefringence from early dark energy}",
    eprint = "2209.07804",
    archivePrefix = "arXiv",
    primaryClass = "astro-ph.CO",
    reportNumber = "RESCEU-16/22",
    doi = "10.1103/PhysRevD.107.L041302",
    journal = "Phys. Rev. D",
    volume = "107",
    number = "4",
    pages = "L041302",
    year = "2023"
}

@article{Birefrengence2,
    author = "Eskilt, Johannes R. and Herold, Laura and Komatsu, Eiichiro and Murai, Kai and Namikawa, Toshiya and Naokawa, Fumihiro",
    title = "{Constraint on Early Dark Energy from Isotropic Cosmic Birefringence}",
    eprint = "2303.15369",
    archivePrefix = "arXiv",
    primaryClass = "astro-ph.CO",
    month = "3",
    year = "2023"
}

@article{Planck_VarMe,
    author = "Ade, P. A. R. and others",
    collaboration = "Planck",
    title = "{Planck intermediate results - XXIV. Constraints on variations in fundamental constants}",
    eprint = "1406.7482",
    archivePrefix = "arXiv",
    primaryClass = "astro-ph.CO",
    doi = "10.1051/0004-6361/201424496",
    journal = "Astron. Astrophys.",
    volume = "580",
    pages = "A22",
    year = "2015"
}

@article{VarMe1,
    author = "Uzan, Jean-Philippe",
    title = "{Varying Constants, Gravitation and Cosmology}",
    eprint = "1009.5514",
    archivePrefix = "arXiv",
    primaryClass = "astro-ph.CO",
    doi = "10.12942/lrr-2011-2",
    journal = "Living Rev. Rel.",
    volume = "14",
    pages = "2",
    year = "2011"
}

@article{VarMe2015,
    author = "Hart, Luke and Chluba, Jens",
    title = "{New constraints on time-dependent variations of fundamental constants using Planck data}",
    eprint = "1705.03925",
    archivePrefix = "arXiv",
    primaryClass = "astro-ph.CO",
    doi = "10.1093/mnras/stx2783",
    journal = "Mon. Not. Roy. Astron. Soc.",
    volume = "474",
    number = "2",
    pages = "1850--1861",
    year = "2018"
}

@article{VarMe2018,
    author = "Hart, Luke and Chluba, Jens",
    title = "{Updated fundamental constant constraints from Planck 2018 data and possible relations to the Hubble tension}",
    eprint = "1912.03986",
    archivePrefix = "arXiv",
    primaryClass = "astro-ph.CO",
    doi = "10.1093/mnras/staa412",
    journal = "Mon. Not. Roy. Astron. Soc.",
    volume = "493",
    number = "3",
    pages = "3255--3263",
    year = "2020"
}

@article{Q_MPCL1,
    author = "Raveri, Marco and Doux, Cyrille",
    title = "{Non-Gaussian estimates of tensions in cosmological parameters}",
    eprint = "2105.03324",
    archivePrefix = "arXiv",
    primaryClass = "astro-ph.CO",
    doi = "10.1103/PhysRevD.104.043504",
    journal = "Phys. Rev. D",
    volume = "104",
    number = "4",
    pages = "043504",
    year = "2021"
}

@ARTICLE{BOBYQA1,
       author = {{Cartis}, Coralia and {Fiala}, Jan and {Marteau}, Benjamin and {Roberts}, Lindon},
        title = "{Improving the Flexibility and Robustness of Model-Based Derivative-Free Optimization Solvers}",
      journal = {arXiv e-prints},
         year = 2018,
        month = mar,
          eid = {arXiv:1804.00154},
        pages = {arXiv:1804.00154},
          doi = {10.48550/arXiv.1804.00154},
archivePrefix = {arXiv},
       eprint = {1804.00154},
 primaryClass = {math.OC},
       adsurl = {https://ui.adsabs.harvard.edu/abs/2018arXiv180400154C}
}

@ARTICLE{BOBYQA2,
       author = {{Cartis}, Coralia and {Roberts}, Lindon and {Sheridan-Methven}, Oliver},
        title = "{Escaping local minima with derivative-free methods: a numerical investigation}",
      journal = {arXiv e-prints},
         year = 2018,
        month = dec,
          eid = {arXiv:1812.11343},
        pages = {arXiv:1812.11343},
          doi = {10.48550/arXiv.1812.11343},
archivePrefix = {arXiv},
       eprint = {1812.11343},
 primaryClass = {math.OC},
       adsurl = {https://ui.adsabs.harvard.edu/abs/2018arXiv181211343C}
}

@ARTICLE{AIC1,
  author={Akaike, H.},
  journal={IEEE Transactions on Automatic Control}, 
  title={A new look at the statistical model identification}, 
  year={1974},
  volume={19},
  number={6},
  pages={716-723},
  doi={10.1109/TAC.1974.1100705}}

@article{Cosmology_intertwined,
    author = "Abdalla, Elcio and others",
    title = "{Cosmology intertwined: A review of the particle physics, astrophysics, and cosmology associated with the cosmological tensions and anomalies}",
    eprint = "2203.06142",
    archivePrefix = "arXiv",
    primaryClass = "astro-ph.CO",
    reportNumber = "FERMILAB-CONF-22-192-SCD",
    doi = "10.1016/j.jheap.2022.04.002",
    journal = "JHEAp",
    volume = "34",
    pages = "49--211",
    year = "2022"
}

@article{Ups_Downs_EDE,
    author = "Poulin, Vivian and Smith, Tristan L. and Karwal, Tanvi",
    title = "{The Ups and Downs of Early Dark Energy solutions to the Hubble tension: a review of models, hints and constraints circa 2023}",
    eprint = "2302.09032",
    archivePrefix = "arXiv",
    primaryClass = "astro-ph.CO",
    month = "2",
    year = "2023"
}

@article{EDE_Silvia_Lennart,
    author = "Smith, Tristan L. and Lucca, Matteo and Poulin, Vivian and Abellan, Guillermo F. and Balkenhol, Lennart and Benabed, Karim and Galli, Silvia and Murgia, Riccardo",
    title = "{Hints of early dark energy in Planck, SPT, and ACT data: New physics or systematics?}",
    eprint = "2202.09379",
    archivePrefix = "arXiv",
    primaryClass = "astro-ph.CO",
    reportNumber = "ULB-TH/22-03, LUPM:22-003",
    doi = "10.1103/PhysRevD.106.043526",
    journal = "Phys. Rev. D",
    volume = "106",
    number = "4",
    pages = "043526",
    year = "2022"
}

@article{EDE_ACT1,
    author = "Hill, J. Colin and others",
    title = "{Atacama Cosmology Telescope: Constraints on prerecombination early dark energy}",
    eprint = "2109.04451",
    archivePrefix = "arXiv",
    primaryClass = "astro-ph.CO",
    doi = "10.1103/PhysRevD.105.123536",
    journal = "Phys. Rev. D",
    volume = "105",
    number = "12",
    pages = "123536",
    year = "2022"
}

@article{Feldman_Profile,
    author = "Feldman, Gary J. and Cousins, Robert D.",
    title = "{A Unified approach to the classical statistical analysis of small signals}",
    eprint = "physics/9711021",
    archivePrefix = "arXiv",
    reportNumber = "HUTP-97-A096",
    doi = "10.1103/PhysRevD.57.3873",
    journal = "Phys. Rev. D",
    volume = "57",
    pages = "3873--3889",
    year = "1998"
}

@article{Tristan_Profile,
    author = "Smith, Tristan L. and Poulin, Vivian and Bernal, Jos{\'e} Luis and Boddy, Kimberly K. and Kamionkowski, Marc and Murgia, Riccardo",
    title = "{Early dark energy is not excluded by current large-scale structure data}",
    eprint = "2009.10740",
    archivePrefix = "arXiv",
    primaryClass = "astro-ph.CO",
    doi = "10.1103/PhysRevD.103.123542",
    journal = "Phys. Rev. D",
    volume = "103",
    number = "12",
    pages = "123542",
    year = "2021"
}

@article{EDE_ACT2,
    author = "Hill, J. Colin and McDonough, Evan and Toomey, Michael W. and Alexander, Stephon",
    title = "{Early dark energy does not restore cosmological concordance}",
    eprint = "2003.07355",
    archivePrefix = "arXiv",
    primaryClass = "astro-ph.CO",
    doi = "10.1103/PhysRevD.102.043507",
    journal = "Phys. Rev. D",
    volume = "102",
    number = "4",
    pages = "043507",
    year = "2020"
}

@article{Concordance_Cosmology,
    author = "Raveri, Marco and Hu, Wayne",
    title = "{Concordance and Discordance in Cosmology}",
    eprint = "1806.04649",
    archivePrefix = "arXiv",
    primaryClass = "astro-ph.CO",
    doi = "10.1103/PhysRevD.99.043506",
    journal = "Phys. Rev. D",
    volume = "99",
    number = "4",
    pages = "043506",
    year = "2019"
}

@article{OLE,
    author = {G{\"u}nther, Sven and Balkenhol, Lennart and Fidler, Christian and Khalife, Ali Rida and Lesgourgues, Julien and Mosbech, Markus R. and Sharma, Ravi Kumar},
    title = "{OL{\'E}-- Online Learning Emulation in Cosmology}",
    eprint = "2503.13183",
    archivePrefix = "arXiv",
    primaryClass = "astro-ph.CO",
    reportNumber = "TTK-25-08",
    month = "3",
    year = "2025"
}

@article{Tristan_Nils,
    author = {Smith, Tristan L. and Sch{\"o}neberg, Nils},
    title = "{Predictions for new physics in the CMB damping tail}",
    eprint = "2503.20002",
    archivePrefix = "arXiv",
    primaryClass = "astro-ph.CO",
    month = "3",
    year = "2025"
}

@article{Gelman_Rubin,
 ISSN = {08834237},
 URL = {http://www.jstor.org/stable/2246093},
 author = {Andrew Gelman and Donald B. Rubin},
 journal = {Statistical Science},
 number = {4},
 pages = {457--472},
 publisher = {Institute of Mathematical Statistics},
 title = {Inference from Iterative Simulation Using Multiple Sequences},
 urldate = {2023-10-27},
 volume = {7},
 year = {1992}
}

@article{SPT_Maps_Paper,
    author = "Quan, W. and others",
    collaboration = "SPT-3G",
    title = "{SPT-3G D1: Maps of the millimeter-wave sky from 2019 and 2020 observations of the SPT-3G Main field}",
    eprint = "2603.20163",
    archivePrefix = "arXiv",
    primaryClass = "astro-ph.CO",
    reportNumber = "FERMILAB-PUB-26-0196-PPD",
    month = "3",
    year = "2026"
}

@article{knox19,
    author = "Knox, Lloyd and Millea, Marius",
    title = "{Hubble constant hunter{\textquoteright}s guide}",
    eprint = "1908.03663",
    archivePrefix = "arXiv",
    primaryClass = "astro-ph.CO",
    doi = "10.1103/PhysRevD.101.043533",
    journal = "Phys. Rev. D",
    volume = "101",
    number = "4",
    pages = "043533",
    year = "2020"
}

@article{CamSpec,
    author = "Rosenberg, Erik and Gratton, Steven and Efstathiou, George",
    title = "{CMB power spectra and cosmological parameters from Planck PR4 with CamSpec}",
    eprint = "2205.10869",
    archivePrefix = "arXiv",
    primaryClass = "astro-ph.CO",
    doi = "10.1093/mnras/stac2744",
    journal = "Mon. Not. Roy. Astron. Soc.",
    volume = "517",
    number = "3",
    pages = "4620--4636",
    year = "2022"
}

@article{Q_MPCL2,
      title="Tensions in cosmology: a discussion of statistical tools to determine inconsistencies",
      author="Matías Leizerovich and Susana J. Landau and Claudia G. Scóccola",
      year="2023",
      eprint="2312.08542",
      archivePrefix="arXiv",
      primaryClass="astro-ph.CO"
}

@article{EDE_Curvature,
    author = "Stevens, Jordan and Khoraminezhad, Hasti and Saito, Shun",
    title = "{Constraining the spatial curvature with cosmic expansion history in a cosmological model with a non-standard sound horizon}",
    eprint = "2212.09804",
    archivePrefix = "arXiv",
    primaryClass = "astro-ph.CO",
    doi = "10.1088/1475-7516/2023/07/046",
    journal = "JCAP",
    volume = "07",
    pages = "046",
    year = "2023"
}

@article{planck_collaboration_planck_2020,
	title = {Planck intermediate results - {LVII}. {Joint} {Planck} {LFI} and {HFI} data processing},
	volume = {643},
	url = {https://doi.org/10.1051/0004-6361/202038073},
	doi = {10.1051/0004-6361/202038073},
	journal = {A\&A},
	author = {{Planck Collaboration} and {Akrami, Y.} and {Andersen, K. J.} and {Ashdown, M.} and {Baccigalupi, C.} and {Ballardini, M.} and {Banday, A. J.} and {Barreiro, R. B.} and {Bartolo, N.} and {Basak, S.} and {Benabed, K.} and {Bernard, J.-P.} and {Bersanelli, M.} and {Bielewicz, P.} and {Bond, J. R.} and {Borrill, J.} and {Burigana, C.} and {Butler, R. C.} and {Calabrese, E.} and {Casaponsa, B.} and {Chiang, H. C.} and {Colombo, L. P. L.} and {Combet, C.} and {Crill, B. P.} and {Cuttaia, F.} and {de Bernardis, P.} and {de Rosa, A.} and {de Zotti, G.} and {Delabrouille, J.} and {Di Valentino, E.} and {Diego, J. M.} and {Doré, O.} and {Douspis, M.} and {Dupac, X.} and {Eriksen, H. K.} and {Fernandez-Cobos, R.} and {Finelli, F.} and {Frailis, M.} and {Fraisse, A. A.} and {Franceschi, E.} and {Frolov, A.} and {Galeotta, S.} and {Galli, S.} and {Ganga, K.} and {Gerbino, M.} and {Ghosh, T.} and {González-Nuevo, J.} and {Górski, K. M.} and {Gruppuso, A.} and {Gudmundsson, J. E.} and {Handley, W.} and {Helou, G.} and {Herranz, D.} and {Hildebrandt, S. R.} and {Hivon, E.} and {Huang, Z.} and {Jaffe, A. H.} and {Jones, W. C.} and {Keihänen, E.} and {Keskitalo, R.} and {Kiiveri, K.} and {Kim, J.} and {Kisner, T. S.} and {Krachmalnicoff, N.} and {Kunz, M.} and {Kurki-Suonio, H.} and {Lasenby, A.} and {Lattanzi, M.} and {Lawrence, C. R.} and {Le Jeune, M.} and {Levrier, F.} and {Liguori, M.} and {Lilje, P. B.} and {Lilley, M.} and {Lindholm, V.} and {López-Caniego, M.} and {Lubin, P. M.} and {Macías-Pérez, J. F.} and {Maino, D.} and {Mandolesi, N.} and {Marcos-Caballero, A.} and {Maris, M.} and {Martin, P. G.} and {Martínez-González, E.} and {Matarrese, S.} and {Mauri, N.} and {McEwen, J. D.} and {Meinhold, P. R.} and {Mennella, A.} and {Migliaccio, M.} and {Mitra, S.} and {Molinari, D.} and {Montier, L.} and {Morgante, G.} and {Moss, A.} and {Natoli, P.} and {Paoletti, D.} and {Partridge, B.} and {Patanchon, G.} and {Pearson, D.} and {Pearson, T. J.} and {Perrotta, F.} and {Piacentini, F.} and {Polenta, G.} and {Rachen, J. P.} and {Reinecke, M.} and {Remazeilles, M.} and {Renzi, A.} and {Rocha, G.} and {Rosset, C.} and {Roudier, G.} and {Rubiño-Martín, J. A.} and {Ruiz-Granados, B.} and {Salvati, L.} and {Savelainen, M.} and {Scott, D.} and {Sirignano, C.} and {Sirri, G.} and {Spencer, L. D.} and {Suur-Uski, A.-S.} and {Svalheim, L. T.} and {Tauber, J. A.} and {Tavagnacco, D.} and {Tenti, M.} and {Terenzi, L.} and {Thommesen, H.} and {Toffolatti, L.} and {Tomasi, M.} and {Tristram, M.} and {Trombetti, T.} and {Valiviita, J.} and {Van Tent, B.} and {Vielva, P.} and {Villa, F.} and {Vittorio, N.} and {Wandelt, B. D.} and {Wehus, I. K.} and {Zacchei, A.} and {Zonca, A.}},
	year = {2020},
	pages = {A42},
}

@article{Camphuis_etal,
    author = "Camphuis, E. and others",
    collaboration = "SPT-3G",
    title = "{SPT-3G D1: CMB temperature and polarization power spectra and cosmology from 2019 and 2020 observations of the SPT-3G Main field}",
    eprint = "2506.20707",
    archivePrefix = "arXiv",
    primaryClass = "astro-ph.CO",
    reportNumber = "FERMILAB-PUB-25-0144-PPD",
    month = "6",
    year = "2025"
}

@article{Carron:2022eyg,
    author = "Carron, Julien and Mirmelstein, Mark and Lewis, Antony",
    title = "{CMB lensing from Planck PR4~maps}",
    eprint = "2206.07773",
    archivePrefix = "arXiv",
    primaryClass = "astro-ph.CO",
    doi = "10.1088/1475-7516/2022/09/039",
    journal = "JCAP",
    volume = "09",
    pages = "039",
    year = "2022"
}

@article{Planck_Legacy,
    author = "Aghanim, N. and others",
    collaboration = "Planck",
    title = "{Planck 2018 results. I. Overview and the cosmological legacy of Planck}",
    eprint = "1807.06205",
    archivePrefix = "arXiv",
    primaryClass = "astro-ph.CO",
    doi = "10.1051/0004-6361/201833880",
    journal = "Astron. Astrophys.",
    volume = "641",
    pages = "A1",
    year = "2020"
}

@article{ACTDR6_main,
    author = "Louis, Thibaut and others",
    collaboration = "ACT",
    title = "{The Atacama Cosmology Telescope: DR6 Power Spectra, Likelihoods and $\Lambda$CDM Parameters}",
    eprint = "2503.14452",
    archivePrefix = "arXiv",
    primaryClass = "astro-ph.CO",
    reportNumber = "FERMILAB-PUB-25-0071-PPD",
    month = "3",
    year = "2025"
}

@article{ACTDR6_Extended,
    author = "Calabrese, Erminia and others",
    collaboration = "ACT",
    title = "{The Atacama Cosmology Telescope: DR6 Constraints on Extended Cosmological Models}",
    eprint = "2503.14454",
    archivePrefix = "arXiv",
    primaryClass = "astro-ph.CO",
    reportNumber = "FERMILAB-PUB-25-0157-PPD",
    month = "3",
    year = "2025"
}

@article{ACTDR6_Lensing,
    author = "Qu, Frank J. and others",
    collaboration = "ACT",
    title = "{The Atacama Cosmology Telescope: A Measurement of the DR6 CMB Lensing Power Spectrum and Its Implications for Structure Growth}",
    eprint = "2304.05202",
    archivePrefix = "arXiv",
    primaryClass = "astro-ph.CO",
    reportNumber = "FERMILAB-PUB-23-237-PPD, FERMILAB-PUB-23-237-PPD",
    doi = "10.3847/1538-4357/acfe06",
    journal = "Astrophys. J.",
    volume = "962",
    number = "2",
    pages = "112",
    year = "2024"
}

@article{desi2025,
    author = "Abdul Karim, M. and others",
    collaboration = "DESI",
    title = "{DESI DR2 Results II: Measurements of Baryon Acoustic Oscillations and Cosmological Constraints}",
    eprint = "2503.14738",
    archivePrefix = "arXiv",
    primaryClass = "astro-ph.CO",
    reportNumber = "FERMILAB-PUB-25-0169-PPD",
    month = "3",
    year = "2025"
}

@article{ACT_DR6_1,
    author = "Louis, Thibaut and others",
    collaboration = "ACT",
    title = "{The Atacama Cosmology Telescope: DR6 Power Spectra, Likelihoods and $\Lambda$CDM Parameters}",
    eprint = "2503.14452",
    archivePrefix = "arXiv",
    primaryClass = "astro-ph.CO",
    reportNumber = "FERMILAB-PUB-25-0071-PPD",
    month = "3",
    year = "2025"
}

@article{ACTDR6_Maps,
    author = "Naess, Sigurd and others",
    collaboration = "ACT",
    title = "{The Atacama Cosmology Telescope: DR6 Maps}",
    eprint = "2503.14451",
    archivePrefix = "arXiv",
    primaryClass = "astro-ph.CO",
    reportNumber = "FERMILAB-PUB-25-0160-PPD",
    month = "3",
    year = "2025"
}

@article{Prabhu_etal,
    author = "Prabhu, K. and others",
    collaboration = "SPT-3G",
    title = "{Testing the \ensuremath{\Lambda}CDM Cosmological Model with Forthcoming Measurements of the Cosmic Microwave Background with SPT-3G}",
    eprint = "2403.17925",
    archivePrefix = "arXiv",
    primaryClass = "astro-ph.CO",
    reportNumber = "FERMILAB-PUB-24-0175-PPD",
    doi = "10.3847/1538-4357/ad5ff1",
    journal = "Astrophys. J.",
    volume = "973",
    number = "1",
    pages = "4",
    year = "2024"
}

@article{Breuval:2024lsv,
    author = "Breuval, Louise and Riess, Adam G. and Casertano, Stefano and Yuan, Wenlong and Macri, Lucas M. and Romaniello, Martino and Murakami, Yukei S. and Scolnic, Daniel and Anand, Gagandeep S. and Soszy\'nski, Igor",
    title = "{Small Magellanic Cloud Cepheids Observed with the Hubble Space Telescope Provide a New Anchor for the SH0ES Distance Ladder}",
    eprint = "2404.08038",
    archivePrefix = "arXiv",
    primaryClass = "astro-ph.CO",
    doi = "10.3847/1538-4357/ad630e",
    journal = "Astrophys. J.",
    volume = "973",
    number = "1",
    pages = "30",
    year = "2024"
}

@article{MUSE,
    author = "Ge, F. and others",
    collaboration = "SPT-3G",
    title = "{Cosmology from CMB lensing and delensed EE power spectra using 2019\textendash{}2020 SPT-3G polarization data}",
    eprint = "2411.06000",
    archivePrefix = "arXiv",
    primaryClass = "astro-ph.CO",
    reportNumber = "FERMILAB-PUB-24-0840-PPD",
    doi = "10.1103/PhysRevD.111.083534",
    journal = "Phys. Rev. D",
    volume = "111",
    number = "8",
    pages = "083534",
    year = "2025"
}

@article{PR4lens,
    author = "Carron, Julien and Mirmelstein, Mark and Lewis, Antony",
    title = "{CMB lensing from Planck PR4~maps}",
    eprint = "2206.07773",
    archivePrefix = "arXiv",
    primaryClass = "astro-ph.CO",
    doi = "10.1088/1475-7516/2022/09/039",
    journal = "JCAP",
    volume = "09",
    pages = "039",
    year = "2022"
}

@article{Khalife:2023qbu,
    author = {Khalife, Ali Rida and Zanjani, Maryam Bahrami and Galli, Silvia and G\"unther, Sven and Lesgourgues, Julien and Benabed, Karim},
    title = "{Review of Hubble tension solutions with new SH0ES and SPT-3G data}",
    eprint = "2312.09814",
    archivePrefix = "arXiv",
    primaryClass = "astro-ph.CO",
    reportNumber = "TTK-23-36",
    doi = "10.1088/1475-7516/2024/04/059",
    journal = "JCAP",
    volume = "04",
    pages = "059",
    year = "2024"
}

@article{AxiEDE_String1,
    author = "McDonough, Evan and Scalisi, Marco",
    title = "{Towards Early Dark Energy in string theory}",
    eprint = "2209.00011",
    archivePrefix = "arXiv",
    primaryClass = "hep-th",
    reportNumber = "MPP-2022-112",
    doi = "10.1007/JHEP10(2023)118",
    journal = "JHEP",
    volume = "10",
    pages = "118",
    year = "2023"
}

@article{AxiEDE_String2,
    author = "Cicoli, Michele and Licheri, Matteo and Mahanta, Ratul and McDonough, Evan and Pedro, Francisco G. and Scalisi, Marco",
    title = "{Early Dark Energy in Type IIB String Theory}",
    eprint = "2303.03414",
    archivePrefix = "arXiv",
    primaryClass = "hep-th",
    reportNumber = "MPP-2023-31",
    doi = "10.1007/JHEP06(2023)052",
    journal = "JHEP",
    volume = "06",
    pages = "052",
    year = "2023"
}

@article{BDrag1,
    author = "Eisenstein, Daniel J. and Hu, Wayne",
    title = "{Baryonic features in the matter transfer function}",
    eprint = "astro-ph/9709112",
    archivePrefix = "arXiv",
    reportNumber = "IASSNS-AST-97-51",
    doi = "10.1086/305424",
    journal = "Astrophys. J.",
    volume = "496",
    pages = "605",
    year = "1998"
}

@article{Bdrag2,
    author = "Hu, Wayne and Sugiyama, Naoshi",
    title = "{Small scale cosmological perturbations: An Analytic approach}",
    eprint = "astro-ph/9510117",
    archivePrefix = "arXiv",
    reportNumber = "IASSNS-AST-95-42, CFPA-TH-95-18, UTAP-212",
    doi = "10.1086/177989",
    journal = "Astrophys. J.",
    volume = "471",
    pages = "542--570",
    year = "1996"
}

@article{candl,
    author = "Balkenhol, L. and Trendafilova, C. and Benabed, K. and Galli, S.",
    title = "{candl: cosmic microwave background analysis with a differentiable likelihood}",
    eprint = "2401.13433",
    archivePrefix = "arXiv",
    primaryClass = "astro-ph.CO",
    doi = "10.1051/0004-6361/202449432",
    journal = "Astron. Astrophys.",
    volume = "686",
    pages = "A10",
    year = "2024"
}

@ARTICLE{Sachs-Wolfe,
       author = {{Sachs}, R.~K. and {Wolfe}, A.~M.},
        title = "{Perturbations of a Cosmological Model and Angular Variations of the Microwave Background}",
      journal = "The Astrophysical Journal",
         year = 1967,
        month = jan,
       volume = {147},
        pages = {73},
          doi = {10.1086/148982},
       adsurl = {https://ui.adsabs.harvard.edu/abs/1967ApJ...147...73S}
}

@article{Vivian_lates_EDE,
    author = "Poulin, Vivian and Smith, Tristan L. and Calder{\'o}n, Rodrigo and Simon, Th{\'e}o",
    title = "{Impact of ACT DR6 and DESI DR2 for Early Dark Energy and the Hubble tension}",
    eprint = "2505.08051",
    archivePrefix = "arXiv",
    primaryClass = "astro-ph.CO",
    month = "5",
    year = "2025"
}

@article{Pantplus1,
    author = "Brout, Dillon and others",
    title = "{The Pantheon+ Analysis: Cosmological Constraints}",
    eprint = "2202.04077",
    archivePrefix = "arXiv",
    primaryClass = "astro-ph.CO",
    doi = "10.3847/1538-4357/ac8e04",
    journal = "Astrophys. J.",
    volume = "938",
    number = "2",
    pages = "110",
    year = "2022"
}

@article{Pantplus2,
    author = "Scolnic, Dan and others",
    title = "{The Pantheon+ Analysis: The Full Data Set and Light-curve Release}",
    eprint = "2112.03863",
    archivePrefix = "arXiv",
    primaryClass = "astro-ph.CO",
    doi = "10.3847/1538-4357/ac8b7a",
    journal = "Astrophys. J.",
    volume = "938",
    number = "2",
    pages = "113",
    year = "2022"
}

@article{ACT_Lensing2,
    author = "Madhavacheril, Mathew S. and others",
    collaboration = "ACT",
    title = "{The Atacama Cosmology Telescope: DR6 Gravitational Lensing Map and Cosmological Parameters}",
    eprint = "2304.05203",
    archivePrefix = "arXiv",
    primaryClass = "astro-ph.CO",
    reportNumber = "FERMILAB-PUB-23-206-PPD",
    doi = "10.3847/1538-4357/acff5f",
    journal = "Astrophys. J.",
    volume = "962",
    number = "2",
    pages = "113",
    year = "2024"
}

@article{Old_EDE1,
    author = "Doran, Michael and Lilley, Matthew J. and Schwindt, Jan and Wetterich, Christof",
    title = "{Quintessence and the separation of CMB peaks}",
    eprint = "astro-ph/0012139",
    archivePrefix = "arXiv",
    reportNumber = "HD-THEP-00-61",
    doi = "10.1086/322253",
    journal = "Astrophys. J.",
    volume = "559",
    pages = "501--506",
    year = "2001"
}

@article{Old_EDE2,
    author = "Wetterich, Christof",
    title = "{Phenomenological parameterization of quintessence}",
    eprint = "astro-ph/0403289",
    archivePrefix = "arXiv",
    reportNumber = "HD-THEP-04-08",
    doi = "10.1016/j.physletb.2004.05.008",
    journal = "Phys. Lett. B",
    volume = "594",
    pages = "17--22",
    year = "2004"
}

@article{Old_EDE3,
    author = "Doran, Michael and Robbers, Georg",
    title = "{Early dark energy cosmologies}",
    eprint = "astro-ph/0601544",
    archivePrefix = "arXiv",
    reportNumber = "HD-THEP-06-01",
    doi = "10.1088/1475-7516/2006/06/026",
    journal = "JCAP",
    volume = "06",
    pages = "026",
    year = "2006"
}

@article{Bellido_Jeffrey,
    author = "Nesseris, Savvas and Garcia-Bellido, Juan",
    title = "{Is the Jeffreys' scale a reliable tool for Bayesian model comparison in cosmology?}",
    eprint = "1210.7652",
    archivePrefix = "arXiv",
    primaryClass = "astro-ph.CO",
    reportNumber = "IFT-UAM-CSIC-12-95",
    doi = "10.1088/1475-7516/2013/08/036",
    journal = "JCAP",
    volume = "08",
    pages = "036",
    year = "2013"
}

@article{SimonsObservatory:2025wwn,
    author = "Abitbol, M. and others",
    collaboration = "Simons Observatory",
    title = "{The Simons Observatory: Science Goals and Forecasts for the Enhanced Large Aperture Telescope}",
    eprint = "2503.00636",
    archivePrefix = "arXiv",
    primaryClass = "astro-ph.IM",
    reportNumber = "FERMILAB-PUB-25-0188-PPD",
    month = "3",
    year = "2025"
}

@article{hu97,
    author = "Hu, Wayne and White, Martin J.",
    title = "{The Damping tail of CMB anisotropies}",
    eprint = "astro-ph/9609079",
    archivePrefix = "arXiv",
    reportNumber = "IASSNS-AST-96-47",
    doi = "10.1086/303928",
    journal = "Astrophys. J.",
    volume = "479",
    pages = "568",
    year = "1997"
}

@ARTICLE{numpy,
  author={van der Walt, Stefan and Colbert, S. Chris and Varoquaux, Gael},
  journal={Computing in Science \& Engineering}, 
  title={The NumPy Array: A Structure for Efficient Numerical Computation}, 
  year={2011},
  volume={13},
  number={2},
  pages={22-30},
  doi={10.1109/MCSE.2011.37}}

@ARTICLE{scipy,
	author = {{Virtanen}, Pauli and {Gommers}, Ralf and {Oliphant},
	Travis E. and {Haberland}, Matt and {Reddy}, Tyler and
	{Cournapeau}, David and {Burovski}, Evgeni and {Peterson}, Pearu
	and {Weckesser}, Warren and {Bright}, Jonathan and {van der Walt},
	St{\'e}fan J.  and {Brett}, Matthew and {Wilson}, Joshua and
	{Jarrod Millman}, K.  and {Mayorov}, Nikolay and {Nelson}, Andrew
	R.~J. and {Jones}, Eric and {Kern}, Robert and {Larson}, Eric and
	{Carey}, CJ and {Polat}, {\.I}lhan and {Feng}, Yu and {Moore},
	Eric W. and {Vand erPlas}, Jake and {Laxalde}, Denis and
	{Perktold}, Josef and {Cimrman}, Robert and {Henriksen}, Ian and
	{Quintero}, E.~A. and {Harris}, Charles R and {Archibald}, Anne M.
	and {Ribeiro}, Ant{\^o}nio H. and {Pedregosa}, Fabian and
	{van Mulbregt}, Paul and {Contributors}, SciPy 1. 0},
	title = "{SciPy 1.0: Fundamental Algorithms for Scientific
	Computing in Python}",
	journal = {Nature Methods},
	year = "2020",
	volume={17},
	pages={261--272},
	adsurl = {https://rdcu.be/b08Wh},
	doi = {https://doi.org/10.1038/s41592-019-0686-2},
}

@ARTICLE{matplotlib,
	author = {{Hunter}, John D.},
	title = "{Matplotlib: A 2D Graphics Environment}",
	journal = {Computing in Science and Engineering},
	year = 2007,
	month = may,
	volume = {9},
	number = {3},
	pages = {90-95},
	doi = {10.1109/MCSE.2007.55},
	adsurl = {https://ui.adsabs.harvard.edu/abs/2007CSE.....9...90H}
}

@article{LaPosta:2021pgm,
    author = "La Posta, Adrien and Louis, Thibaut and Garrido, Xavier and Hill, J. Colin",
    title = "{Constraints on prerecombination early dark energy from SPT-3G public data}",
    eprint = "2112.10754",
    archivePrefix = "arXiv",
    primaryClass = "astro-ph.CO",
    doi = "10.1103/PhysRevD.105.083519",
    journal = "Phys. Rev. D",
    volume = "105",
    number = "8",
    pages = "083519",
    year = "2022"
}

@article{Metropolis_Hastings1,
 ISSN = {00063444},
 URL = {http://www.jstor.org/stable/2334940},
 author = {W. K. Hastings},
 journal = {Biometrika},
 number = {1},
 pages = {97--109},
 publisher = {[Oxford University Press, Biometrika Trust]},
 title = {Monte Carlo Sampling Methods Using Markov Chains and Their Applications},
 urldate = {2023-10-27},
 volume = {57},
 year = {1970}
}

@ARTICLE{Metropolis_Hastings2,
       author = {{Metropolis}, Nicholas and {Rosenbluth}, Arianna W. and {Rosenbluth}, Marshall N. and {Teller}, Augusta H. and {Teller}, Edward},
        title = "{Equation of State Calculations by Fast Computing Machines}",
      journal = {Journal of Chemical Physics},
         year = 1953,
        month = jun,
       volume = {21},
       number = {6},
        pages = {1087-1092},
          doi = {10.1063/1.1699114},
       adsurl = {https://ui.adsabs.harvard.edu/abs/1953JChPh..21.1087M}
}

@article{Jhaveri:2026bla,
    author = "Jhaveri, Tanisha and Karwal, Tanvi and Crawford, Thomas and Hu, Wayne and Khalife, Ali Rida and Balkenhol, Lennart and Ge, Fei",
    title = "{Disentangling cosmic distance tensions with early and late dark energy}",
    eprint = "2604.08530",
    archivePrefix = "arXiv",
    primaryClass = "astro-ph.CO",
    month = "4",
    year = "2026"
}
\end{document}